\documentclass[aps,reprint,amsmath,amssymb,floatfix]{revtex4-2}

\usepackage[utf8]{inputenc}
\usepackage[T1]{fontenc}
\usepackage{graphicx}
\usepackage[dvipsnames]{xcolor}
\usepackage{bm}
\usepackage{mathptmx}
\usepackage{etoolbox}
\usepackage[colorlinks=true,
            linkcolor=blue,
            citecolor=blue,
            urlcolor=blue]{hyperref}

\begin{document}

\title{Glassy dynamics, crossover temperature and density scaling in fragile glass-formers}

\author{Ankit Singh}
\email{ankit.singh@unimi.it}
\affiliation{Department of Physics, University of Milan, Italy}
\thanks{A.S. and V.V. contributed equally to this work.} 

\author{Vinay Vaibhav}
\email{vinay.vaibhav@uni-goettingen.de}
\affiliation{Institut für Theoretische Physik, Göttingen, Germany}
\altaffiliation{A.S. and V.V. contributed equally to this work.}

\author{Swarn Lata Singh}
\email{swarnbhu@gmail.com}
\affiliation{Department of Physics, Mahila Mahavidyalaya (MMV), Banaras Hindu University, Varanasi 221005, India}

\author{Yashwant Singh}
\email{singhyas44@gmail.com}
\affiliation{Department of Physics, Banaras Hindu University, Varanasi 221005, India}

\begin{abstract}
We investigate the slowing down of dynamics in a glass-forming mixture interacting via an inverse-power-law (IPL) potential using a combination of theory and large-scale molecular dynamics simulations. We measure the static pair-correlation function, configurational entropy, inherent-structure energy, and structural relaxation time. We employ a theoretical framework to calculate the structural relaxation time $\tau_{\alpha}$, which is found to be in very good agreement with the simulation results. The theory identifies a local structural order which defines the cooperativity of the relaxation and brings forth a fluctuation induced parameter $\psi ( T )$ and a crossover temperature $T_a$ that characterize the density and temperature dependence of the glassy dynamics. Furthermore, we determine a crossover temperature using  independent dynamical and thermodynamic criteria and compare with the theoretically predicted crossover temperature $T_a$. Relaxation dynamics is shown to obey density–temperature scaling, similar to thermodynamic properties, in terms of a variable $\Gamma$  formed by an appropriate combination of density and temperature, characteristic of IPL interactions. Finally, we show that, when the excess thermodynamic and dynamic quantities obtained at different densities are plotted as functions of the reduced temperature $T/T_a$ (or $T_a/T$), the data collapse onto master curves with excellent agreement between theory and simulation. These scaling relations provide a unified description of the thermodynamics and dynamics in IPL systems, enabling the prediction of relaxation behavior over a wide range of densities from data at a single state point.
\end{abstract}                   
\maketitle
\section{Introduction} \label{intro}
The dramatic slowdown of dynamics upon approaching the glass transition is a hallmark of glassy dynamics which manifests in the rapid growth of viscosity and relaxation times, as well as in the decoupling of microscopic and macroscopic time scales \cite{angell1995formation, debenedetti2001supercooled, Mallamace_2010,EdigerHarrowell_2012, berthier2011theoretical}. Understanding how key dynamical quantities depend on control parameters such as temperature, pressure, density, etc., and on interparticle interactions, remains a central challenge, as unlike conventional critical phenomena, the glass transition is not associated with a universally accepted thermodynamic singularity or an easily identifiable diverging correlation length \cite{EdigerHarrowell_2012, berthier2011theoretical, donati1998stringlike, gotze2009complex, BouchaudBiroli2004, MontanariSemerjian2006, Kawasaki2007, KivelsonTarjus2008,Biroli2008, Mosayebi2010, KurchanLevine2011, SaussetLevine2011, tarjus2011overview, Hocky2012, CammarotaBiroli2012, biroli2013perspective, Schoenholz_2016, Berthier2017, NissHecksher2018,
BerthierReichman2023, Berthier_PRX_2026, DyreEdiger2026}. Beginning at least from Adam and Gibbs \cite{Adam} in the mid-1960s, many conjectures and theories based on qualitatively different perspectives have been advanced to uncover the physical mechanism underlying glassy dynamics.  Comprehensive review of these methods can be found in a number of recent articles ~\cite{BouchaudBiroli2004, MontanariSemerjian2006, Kawasaki2007, KivelsonTarjus2008, Biroli2008, Mosayebi2010, KurchanLevine2011, SaussetLevine2011, tarjus2011overview, Hocky2012, CammarotaBiroli2012, biroli2013perspective,Schoenholz_2016, Berthier2017, NissHecksher2018, BerthierReichman2023, Berthier_PRX_2026, DyreEdiger2026}. 

In previous papers ~\cite{PhysRevE.99.030101, PhysRevE.103.032611, PhysRevE.103.052105, PhysRevE.107.014119}, of this series, a theory for slowing down of dynamics in fragile glass-formers was developed. The key step of the theory is to identify and calculate number of particles that form a cluster which acts as a ``cooperatively reorganizing cluster (CRC)'' and its dependence on temperature and density. In a relaxation process, the CRC rearrange irreversibly; the energy involved in the process is the effective activation energy of relaxation. The theory has been found to give a very accurate account of the structural relaxation time $\tau_{\alpha}$ of several model systems belonging to both, the thermal \cite{PhysRevE.103.032611, PhysRevE.103.052105} and athermal \cite{PhysRevE.99.030101, PhysRevE.107.014119} systems. In a thermal system, temperature $T$ is the main control variable and the relaxation time has a crossover at a temperature $T_a$, which separates the high-temperature behaviour from the low-temperature behaviour. In an athermal system temperature is irrelevant apart from rescaling quantities, density (or packing fraction) is the control variable.

The theory identifies a temperature-dependent parameter $\psi(T)$, which measures the effect of fluctuations embedded in the system on stabilizing the size and shape of the CRC \cite{PhysRevE.103.032611}. In athermal systems, $\psi=1$, whereas in thermal systems $\psi$ becomes temperature dependent. In particular, for temperatures above a ``crossover'' temperature $T_a$, one finds $\psi(T)=1$, while for $T<T_a$ the parameter takes a turn and starts decreasing on cooling.  There is a one-to-one correspondence between the crossover region which separates the high temperature behavior from the low temperature behavior of $\psi(T)$ and the relaxation time $\tau_{\alpha}$ indicating the importance of fluctuations on the activated dynamics. Both $\psi(T)$ and $T_{a}$ depend on details of intermolecular interactions. For example, the well-known difference in the relaxation dynamics of the Lennard-Jones (LJ) and Weeks-Chandler-Andersen (WCA) glass-former ~\cite{tong2020role, landes2020attractive, Chattoraj_2020,pedersen2010repulsive, banerjee2014role,Schweizer_2015}, despite their similar pair correlation functions, can be understood in terms of the different values of $T_a$ and $\psi(T)$ \cite{PhysRevE.103.052105}. Though in both systems $T_a$ is found to follow a power-law form $T_a=a_0\rho^{\gamma},$ there is a subtle difference in the nature of the slowing down of dynamics which becomes apparent when the data for $\tau_\alpha$ are plotted as a function of $T_a/T$. While in the case of the LJ system the data for different densities collapse onto a master curve, in the case of the WCA system such a collapse fails~\cite{PhysRevE.103.052105}.

To gain a better understanding of the role played by the intermolecular interactions in slowing down of dynamics and its relation with thermodynamics, we, in this work investigate equilibrium and dynamic properties of a glass-forming liquid interacting via an inverse-power-law (IPL) potential. The form of the IPL potentials (see Eq.~\eqref{potential-n}-\eqref{Gamma}) allows one to combine the two independent variables, density $\rho$ and temperature $T$, into one which we denote by  $\Gamma$ \cite{Hoover,WHoover,SINGH1991351}. The parameter $\Gamma$ provides an additional handle to investigate the scaling properties of the dynamics and their relationship with thermodynamics.

In Sec.~\ref{sec:model_method} we use large-scale molecular-dynamics simulations to calculate static pair-correlation functions, configurational entropy, inherent-structure energy, and the self-intermediate scattering function $F_s(q,t)$ (see Eq.~\eqref{Fsqt} its definition). When derivative of function $F_s(q,t)$ with respect to $\log\, t$ is taken a maximum which separates two minima is found to emerge  below a certain temperature $T_h$~\cite{coslovich2025freezing}. Since, at the temperature $T_h$, the time dependence of the function $F_s(q,t)$ shows a change from its high-temperature behavior, it can be identified as a crossover temperature. A function $\phi(T)=1-h(T)$, where $h(T)$ is the height of the maximum at temperature $T$, is found to have features similar to those of $\psi(T)$ determined theoretically in Sec.~\ref{theory}. The outline of the theory formulated in Refs.~\cite{PhysRevE.99.030101, PhysRevE.103.032611, PhysRevE.103.052105, PhysRevE.107.014119} is given in Sec.~\ref{theory}. In Sec.~\ref{results} we compare the results obtained from the theory with those from the simulations. In Sec.~\ref{scaling}, scaling of the results in terms of $T_a$ and $\Gamma$ is discussed. In Sec.~\ref{conclusion}, we summarize our findings.

\section{Model glass-former and simulations of its static and dynamic properties}\label{sec:model_method}
We investigate a 50:50 binary mixture \cite{bernu1985molecular,bernu1987soft,Ninarello_PhysRevX_2017} $(A:B)$ of $N\,=N_A+N_B$ particles in a volume $V$, interacting via the IPL potential,
\begin{equation}
u_{\alpha\gamma}(r) = \epsilon_{\alpha\gamma} \left(\frac{\sigma_{\alpha\gamma}}{r} \right)^n,
\label{potential-n}
\end{equation}
where $\alpha,\gamma\in\{A,B\}$. Parameters $\epsilon_{\alpha\gamma}$ and $\sigma_{\alpha\gamma}$ set the energy and length scales, respectively, $n$ measures the softness of the interaction, and $r$ is the interparticle separation. Eq.~\eqref{potential-n} can also be written as
\begin{equation}
\beta u_{\alpha\gamma}(r) = \left(\frac{\Gamma_{\alpha\gamma}}{r^3} \right)^{n/3},
\label{eq:scaled-ipl-potential}
\end{equation}
where $\Gamma_{\alpha\gamma} = \rho\sigma_{\alpha\gamma}^{3} \left( \beta\epsilon_{\alpha\gamma} \right)^{3/n}$ and $r$ is measured in units of $(1/\rho)^{1/3}$. Here, $\rho$ is number density and $\beta =({k_{\mathrm B}T})^{-1}$; $k_{\mathrm B}$ is the Boltzmann constant and $T$ is the temperature. For a binary mixture, the combined parameter $\Gamma$ \cite{Hoover,WHoover,SINGH1991351} is defined as
\begin{equation}
\Gamma = x_A^2\Gamma_{AA} + 2x_Ax_B\Gamma_{AB} + x_B^2\Gamma_{BB},
\label{Gamma}
\end{equation}
where $ x_\alpha={N_\alpha/}{N}$. The parameter $\Gamma$ is used in Sec.~\ref{scaling} to scale the static and dynamic quantities.

We use the potential of Eq.~\eqref{potential-n} to calculate the static and dynamic quantities at three densities, $\rho=0.75$, $0.80$, and $0.85$, for several values of $T$. Parameters used in the calculations are $n=12$, $\sigma_{BB}=1.2\sigma_{AA}$, $\sigma_{AB}=1.1\sigma_{AA}$, and $\epsilon_{BB} =\epsilon_{AB}= \epsilon_{AA}$. All physical quantities are expressed in reduced units; length in units of $\sigma_{AA}$, energy in units of $\epsilon_{AA}$, temperature in units of $\epsilon_{AA}/k_{\mathrm B}$, and time in units of $\sqrt{m\sigma_{AA}^{2}/\epsilon_{AA}}$. Furthermore, we set $\sigma_{AA}=1.0$, $\epsilon_{AA}=1.0$, $k_{\mathrm B}=1$, and the mass $m=1$. Particles of both species have identical Maxwell-Boltzmann distributions.

In the simulation study, the potential is set equal to zero for $r\geq r_{\mathrm{cut}}$, where $r_{\mathrm{cut}} = \sqrt{3}\,\sigma_{\alpha\gamma}$. However, to ensure continuity of the potential and its first two derivatives at the cutoff point, we add a biquadratic correction term $v_{\alpha\gamma}(r)$ to the original interaction $u_{\alpha\gamma}(r)$, such that the total interaction becomes $u_{\alpha\gamma}(r) + v_{\alpha\gamma}(r)$. The correction term is given by \cite{Ninarello_PhysRevX_2017}
\begin{equation}\label{smoothEq}
v_{\alpha \gamma}(r) = c_0 + c_2 \left(\frac{r}{\sigma_{\alpha \gamma}}\right)^2 + c_4 \left(\frac{r}{\sigma_{\alpha \gamma}}\right)^4,
\end{equation}
where the coefficients $c_0$, $c_2$, and $c_4$ are chosen to enforce smoothness at $r_{\rm cut}$. Molecular dynamics simulations are performed using the LAMMPS package \cite{thompson2022lammps}, with the equations of motion integrated via the velocity-Verlet algorithm under periodic boundary conditions. The system consists of $N = 10{,}000$ particles. To generate well-equilibrated supercooled configurations, each independent sample is first equilibrated at a high temperature in the liquid state and subsequently quenched to the target temperature. This is followed by further equilibration over sufficiently long times in the NVT ensemble, with temperature controlled using a Nosé–Hoover thermostat. Production runs are then carried out at constant temperature and volume. For each state point (specified by density and temperature), we prepare a total of $16$ independent samples to ensure adequate statistical averaging.

\subsection{Pair correlation functions}\label{sec:structure}
\begin{figure*}[]
\includegraphics[width=0.98\textwidth,clip]{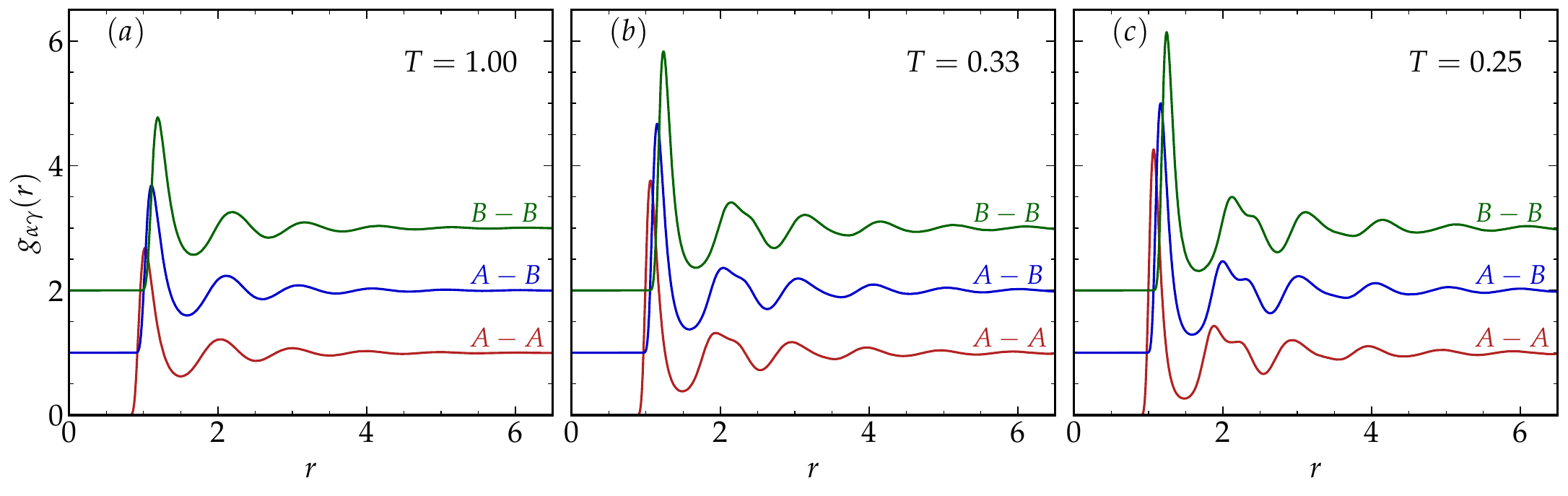}
\caption{Partial radial distribution functions $g_{\alpha\gamma}(r)$ for $A$–$A$, $A$–$B$, and $B$–$B$ correlations at density $\rho = 0.80$ for temperatures (a) $T = 1.00$, (b) $T = 0.33$, and (c) $T = 0.25$. For clarity, the $g_{AB}(r)$ and $g_{BB}(r)$ curves are vertically shifted upward by 1 and 2 units, respectively.}
\label{fig:g_partial}
\end{figure*}

The static order in a liquid is defined in terms of a two-body correlation function, which, in the case of simple fluids, reduces to the radial distribution function \cite{Hansen}. For a binary mixture, the radial distribution function $g(r)$ is given as,
\begin{equation}
g(r)=x_A^2g_{AA}(r)+2x_Ax_Bg_{AB}(r)+x_B^2g_{BB}(r),
\end{equation}
where $g_{\alpha\gamma}(r)$ is the partial radial distribution function. In simulations, $g_{\alpha\gamma}(r)$ is evaluated from the relation, 
\begin{equation}
g_{\alpha\gamma}(r) = \frac{V}{N_{\alpha}N_{\gamma}} \left\langle \sum_{j=1}^{N_{\alpha}} \sum_{\substack{k=1 \\ k \neq j}}^{N_{\gamma}} \delta(\vec{r}-\vec{r}_{j}+\vec{r}_{k}) \right\rangle,
\end{equation}
In Figs.~\ref{fig:g_partial} (a-c), we plot $g_{\alpha\gamma}(r)$ for $\rho=0.8$ at three values of temperature, $T=1.00$ in the normal liquid regime, $T=0.33$ in the moderately supercooled regime below the onset of glassiness, and $T=0.25$ deep in the supercooled state. As temperature decreases, the height of the first and subsequent peaks in all $g_{\alpha\gamma}(r)$ increases, reflecting enhanced local ordering and tighter packing as the system approaches the glass transition. The positions of the peaks remain close to the characteristic interparticle distances set by the size parameters $\sigma_{\alpha\gamma}$, but the sharpening of the peaks signals the growing structural rigidity and the formation of more well-defined coordination shells. An important structural signature observed in Fig.~\ref{fig:g_partial} is the splitting of the second peak of the partial radial distribution functions upon cooling. At the highest temperature $T = 1.00$, the second peak appears as a broad and smooth maximum, reflecting relatively disordered local packing and significant thermal motion. However, as the temperature decreases to $T = 0.33$ and further to $T = 0.25$, this peak clearly splits into two distinct sub-peaks. This splitting phenomenon is widely regarded as a mark of increased medium-range order and the formation of well-defined local coordination structures in supercooled liquids and glassy states \cite{binder2011glassy, berthier2009nonperturbative}, indicating enhanced local packing constraints. The emergence of these sub-peaks highlights the growing structural complexity accompanying dynamic slowdown in the glass-former.

\begin{figure*}[t!]
\includegraphics[width=0.33\textwidth,clip]{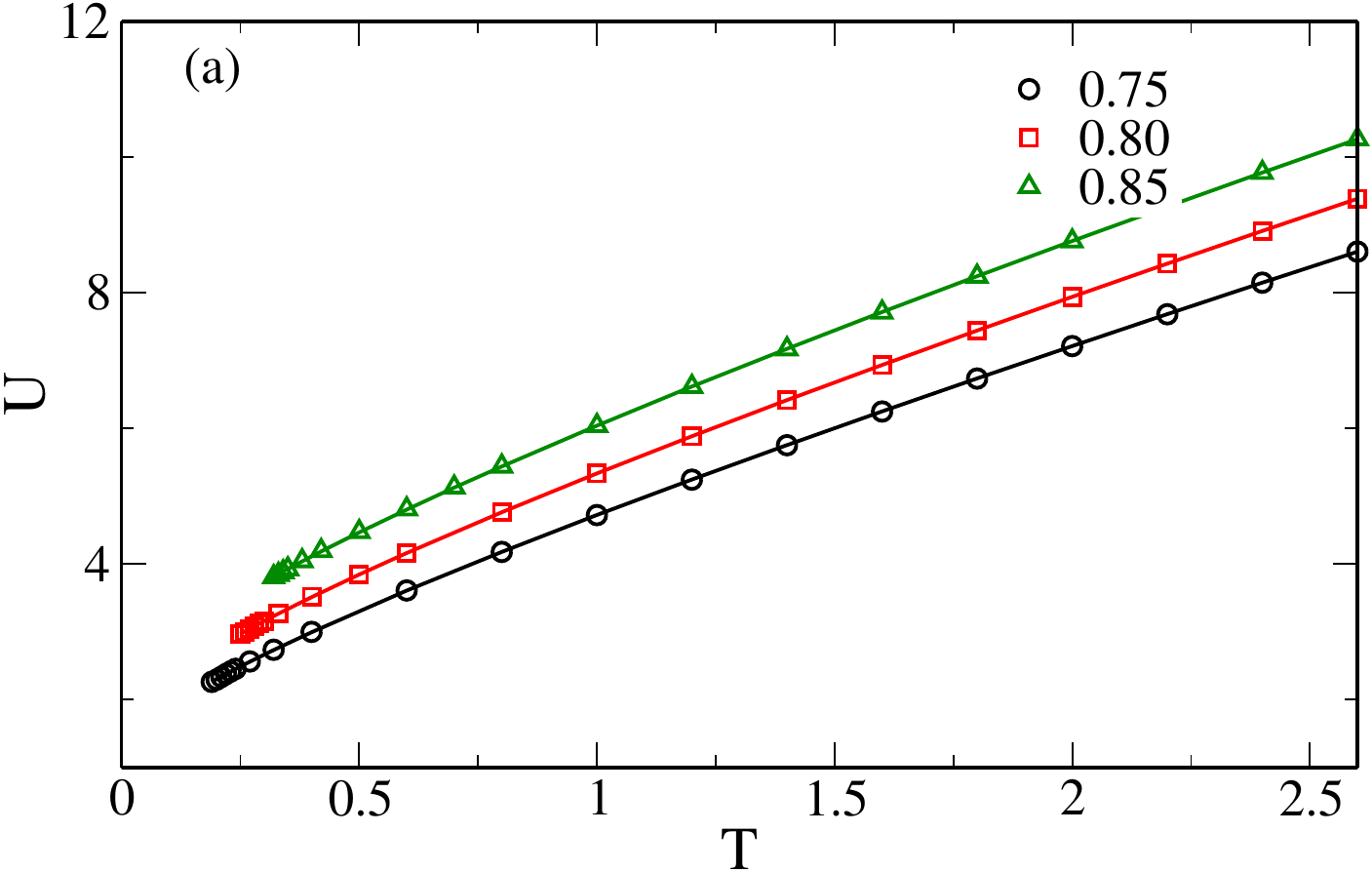}
\includegraphics[width=0.33\textwidth,clip]{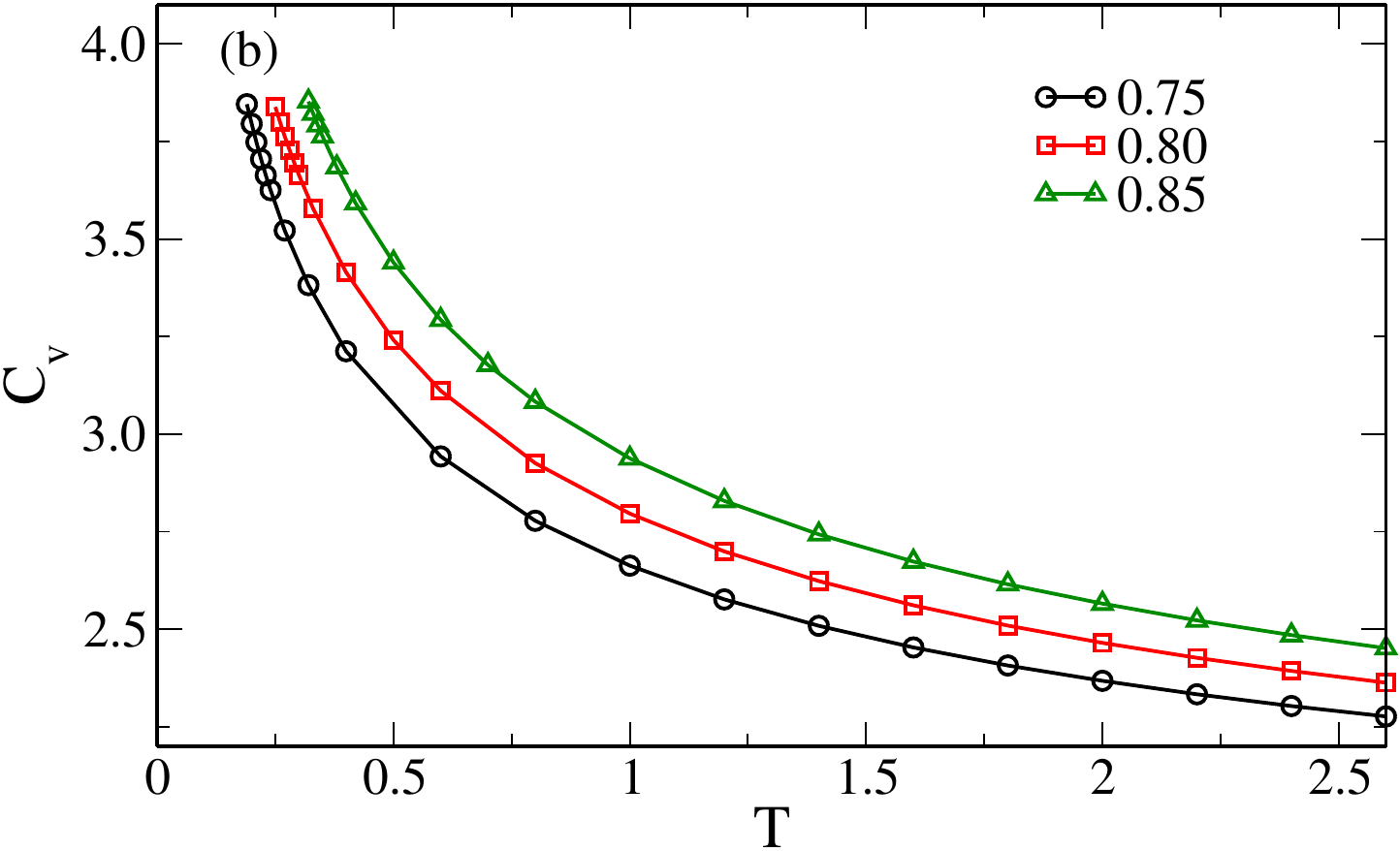}
\includegraphics[width=0.33\textwidth,clip]{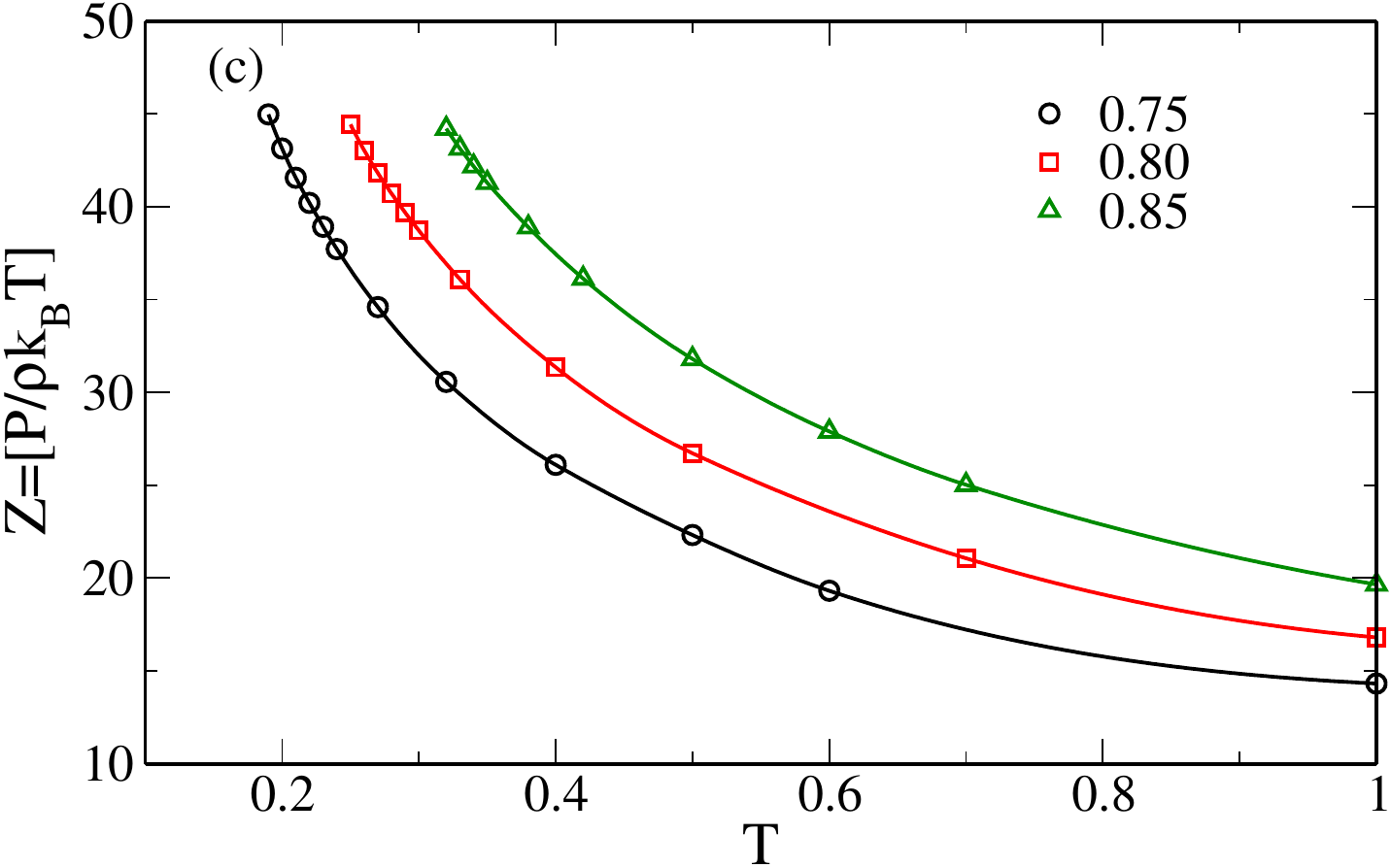}
\caption{(a) Internal energy per particle, $U(T)$, plotted as a function of temperature, $T$, for three densities, $\rho=0.75$, $0.80$, and $0.85$. The symbols represent the calculated values, while the lines show fits to the algebraic relation~\cite{rosenfeld1998density}, $U(T)=\frac{3}{2}T+d_0T^{d_1}+d_2$, where $d_0$ and $d_2$ depend on density and $d_1=0.58$. (b) Specific heat, $C_V(T)$, plotted as a function of $T$ for the same densities. The symbol-lines are calculated using $C_V(T)=\frac{3}{2}+d_0d_1T^{d_1-1}$, obtained from the fitted expression for $U(T)$. (c) Reduced pressure, $Z=\beta P/\rho$, plotted as a function of $T$. The symbols represent the calculated values, while the lines show Akima spline fits.}
\label{U_Cv_Z}
\end{figure*}

The values of $g_{\alpha\gamma}(r)$ are used to calculate the internal energy per particle $U(T)$, the specific heat $C_V(T)$, and the reduced pressure, defined as $Z=\beta P/\rho$. We plot them respectively, in Figs.~\ref{U_Cv_Z}~(a-c) for the three densities as functions of $T$. Values of $U(T)$ and $C_V(T)$ are found to satisfy the algebraic relations given as \cite{rosenfeld1998density}, $U(T) = \frac{3}{2}T+d_0T^{d_1}+d_2$ and $C_V(T)=\frac{3}{2}+d_0d_1T^{d_1-1}$. Here, $d_1=0.58$, while the coefficients $d_0$ and $d_2$ depend on density. For $\rho=0.75$, $d_0=2.00$ and $d_2=1.21$; for $\rho=0.80$, $d_0=2.23$ and $d_2=1.60$; and for $\rho=0.85$, $d_0=2.47$ and $d_2=2.05$. From Fig.~\ref{U_Cv_Z}, we observe that the internal energy $U$ decreases upon lowering the temperature at a constant density, while at a given temperature, the internal energy increases with increasing density because of an increase repulsive interactions between particles. We also observe that the specific heat, $C_v$, and the reduced pressure, $Z$, increase upon lowering the temperature at a fixed density and increase with density at a constant temperature.

\subsection{Configurational entropy and inherent structure energy}\label{entropy}

\begin{figure}[b]
\includegraphics[width=0.39\textwidth,clip]{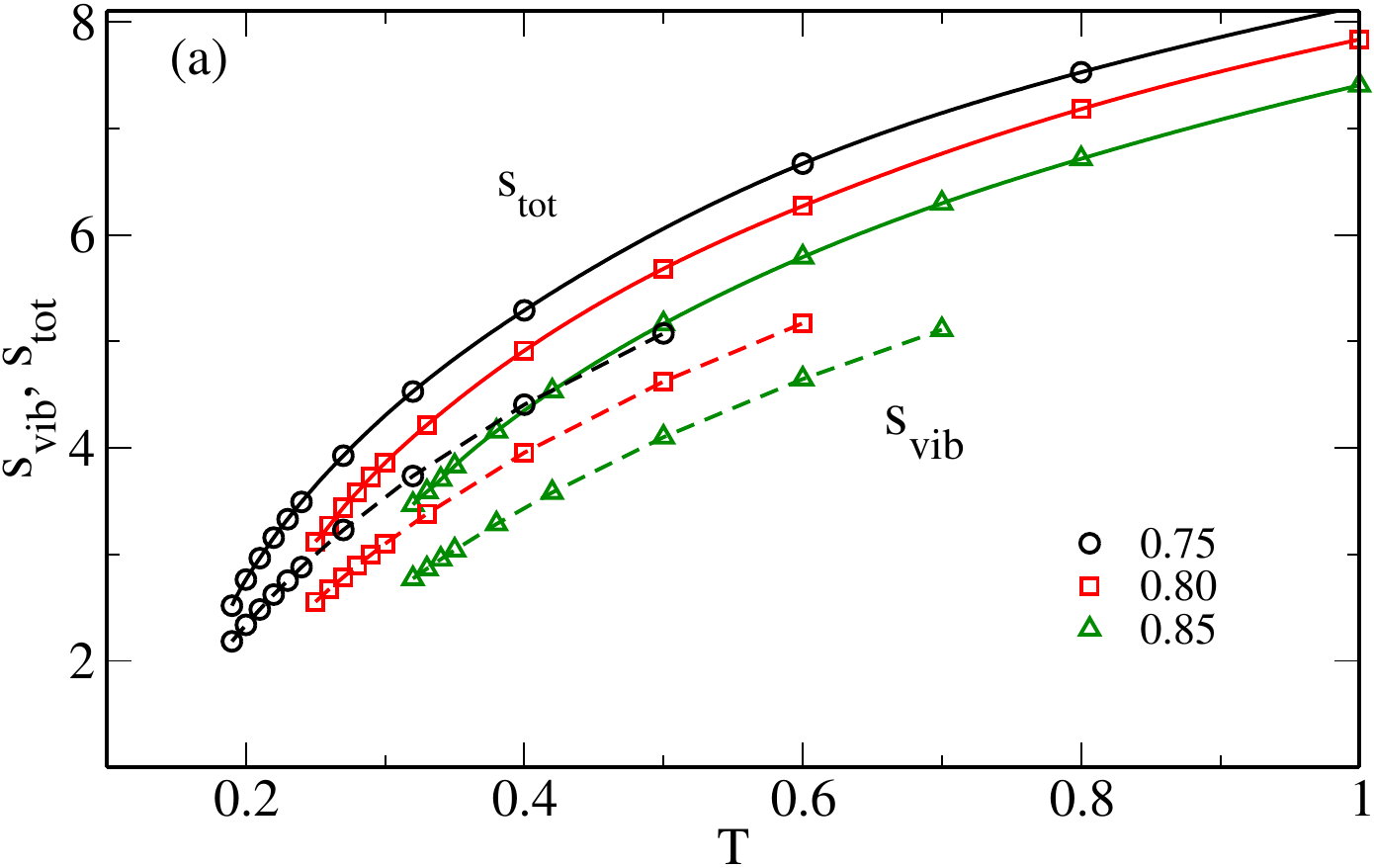}
\includegraphics[width=0.4\textwidth,clip]{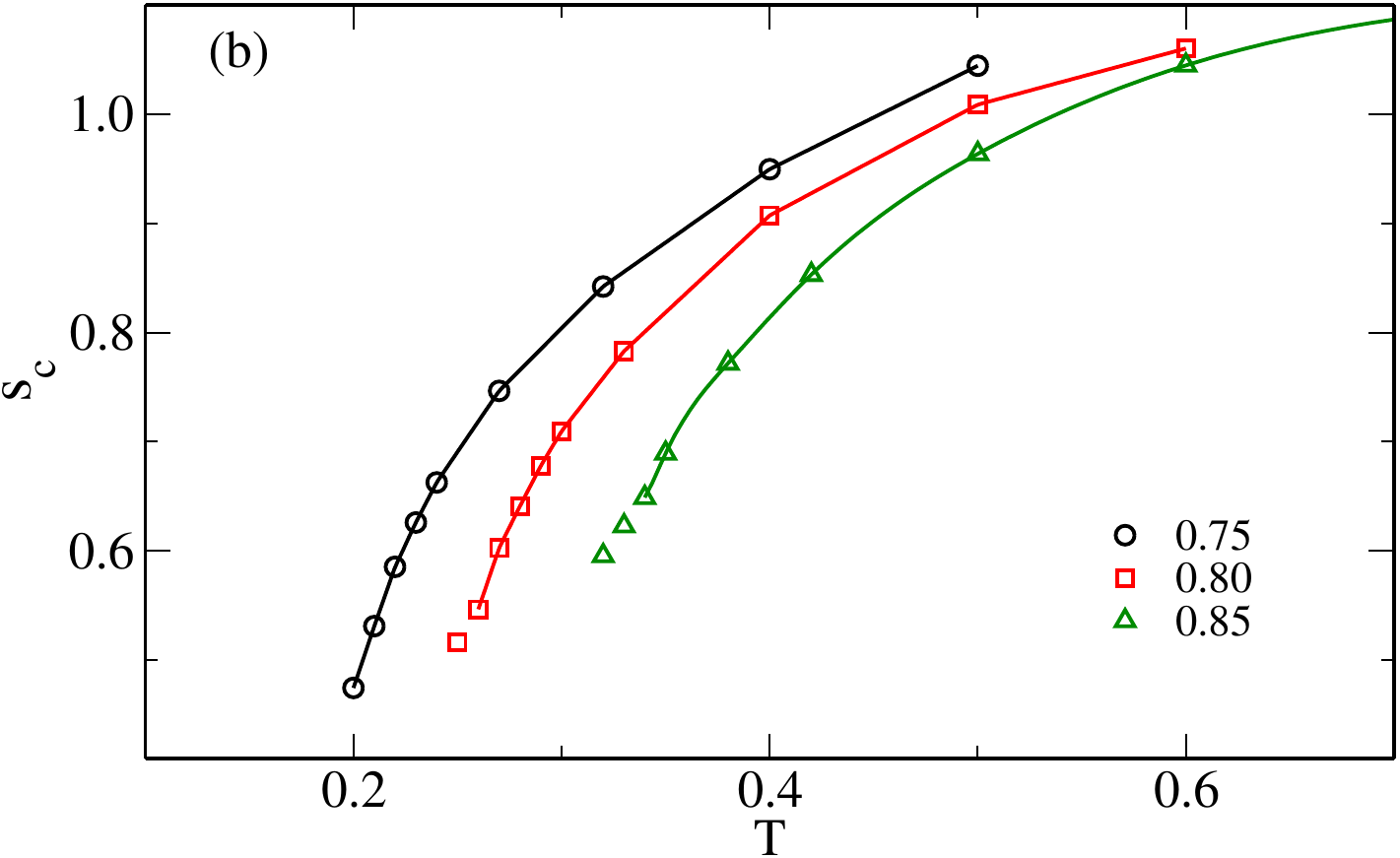}
\caption{Temperature dependence of the total entropy per particle $s_{\mathrm{tot}}=S_{\mathrm{tot}}/N$ (solid lines with open symbols) and vibrational entropy per particle $s_{\mathrm{vib}}=S_{\mathrm{vib}}/N$ (dashed lines with open symbols) for densities $\rho = 0.75$, $0.80$, and $0.85$ (as indicated) in (a). The corresponding configurational entropy per particle $s_{\mathrm{c}}$ as a function of temperature $T$ in (b).}
\label{Sc}
\end{figure}

The configurational entropy, $S_{\rm conf}$, is obtained by subtracting the vibrational entropy, $S_{\rm vib}$, from the total entropy, $S_{\rm tot}$ \cite{sastry2000evaluation, berthier2019configurational},
\begin{equation}
S_{\rm conf}(T)=S_{\rm tot}(T)-S_{\rm vib}(T).
\label{eq:conf_ent}
\end{equation}
Here, $S_{\rm tot}$ accounts for all possible microstates of the supercooled liquid, $S_{\rm vib}$ accounts for the microstates of the frozen structure (glass) arising due to small vibrational motions in the neighbourhood of the structure. 

The total entropy can be determined via thermodynamic integration from a known reference state \cite{sastry2000evaluation, berthier2019configurational}. We first define the ideal-gas entropy $S_{\mathrm{id}}$ for a binary mixture,
\begin{equation}
S_{\mathrm{id}}/N =  \left[\frac{5}{2} - \ln \rho - \ln \Lambda^3 \right] + S_{\mathrm{mix}}/N,
\end{equation}
where $\Lambda = \sqrt{\beta h^2 / (2\pi m)}$ is the thermal de Broglie wavelength and the mixing entropy contribution (under Stirling's approximation) can be written as $S_{\mathrm{mix}} = -N \left( X_A \ln X_A + X_B \ln X_B \right)$. To find the entropy of the liquid, the integration is performed in two primary stages. At first a density integration is performed to calculate excess entropy $S_{\mathrm{ex}}$ at a high reference temperature $T_r$ (here we choose $T_r = 3$), where the system behaves nearly as an ideal-gas in the low density limit, by integrating over density,
\begin{equation}
S_{\mathrm{ex}}(\rho, T_r)/N = \frac{1}{T_r} \int_{0}^{\rho} \frac{d\rho'}{\rho'} \left( \frac{P(\rho', T_r)}{\rho'} - T_r \right).
\end{equation}
In practice, a small finite lower bound on density (here, 0.005) is used to avoid numerical instabilities at the limit $\rho \to 0$, with the integral evaluated via Gauss quadrature.

The second step is a temperature integration. The total entropy per particle at the target temperature $T$ is then obtained by integrating the averaged potential energy (per particle) $U_p=U-\frac{3}{2}T$ from $T_r$ down to $T$,
\begin{equation}
    \frac{S_{\mathrm{tot}}(T)}{N} = \frac{S_{\mathrm{id}} + S_{\mathrm{ex}}(\rho, T_r)}{N} + \frac{U_p(\rho,T)}{T} - \int_T^{T_{r}} \frac{U_p(\rho,T')}{T'^2} dT'.
    \label{eq:s_tot}
\end{equation}

\begin{figure}[b]
\includegraphics[width=0.4\textwidth,clip]{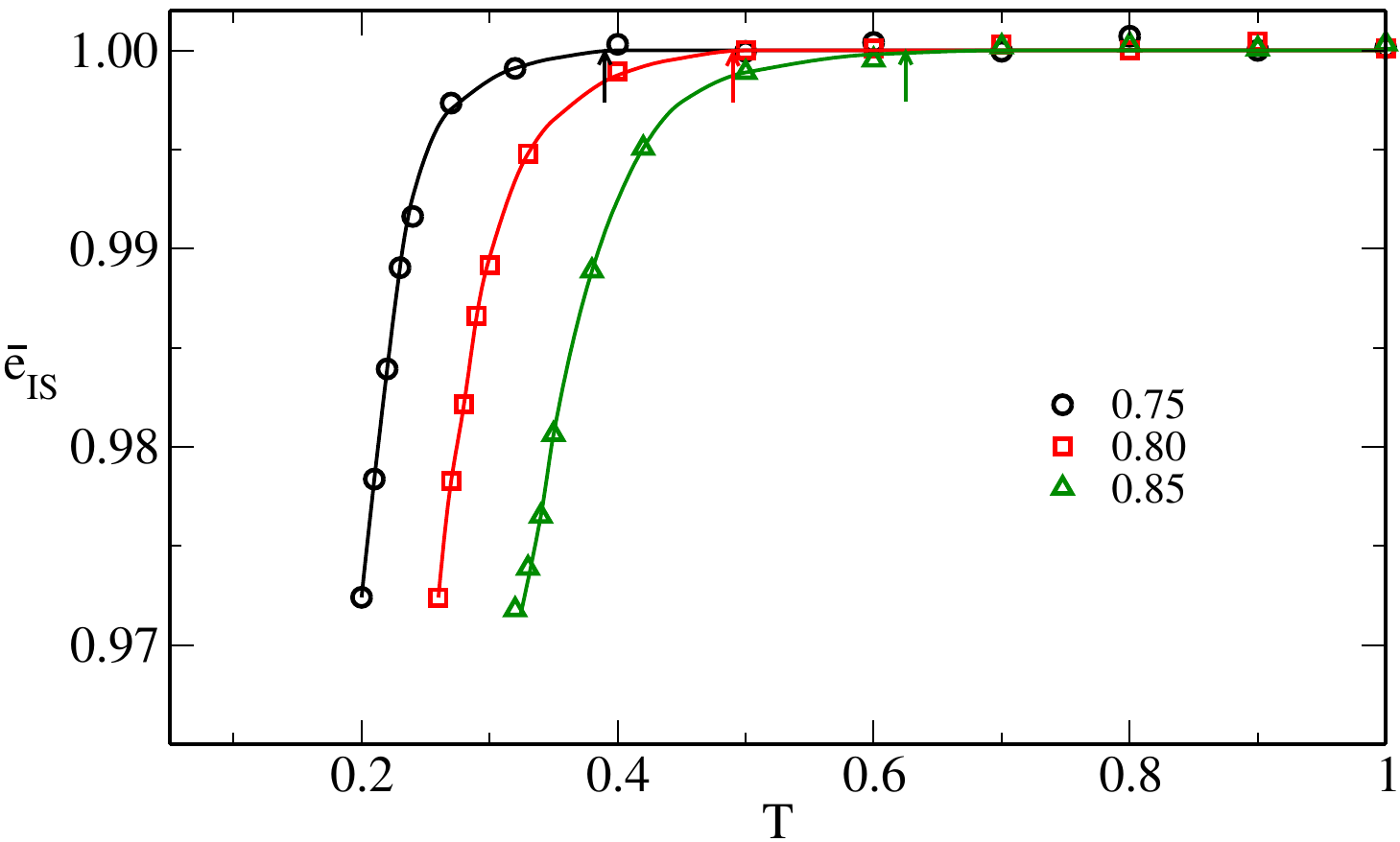}
\caption{Normalized inherent structure energy $\bar{e}_{\rm IS}$ plotted as function of temperature $T$ for the densities $\rho = 0.75, 0.80,$ and $0.85$. Symbols represent the calculated values, while the solid lines represent the corresponding least square fit. The arrow indicates the crossover temperature $T_{IS}$ for different densities, which are listed in Table \ref{table}.}
\label{IS}
\end{figure}

\begin{figure*}[htb]
\includegraphics[width=0.9\textwidth,clip]{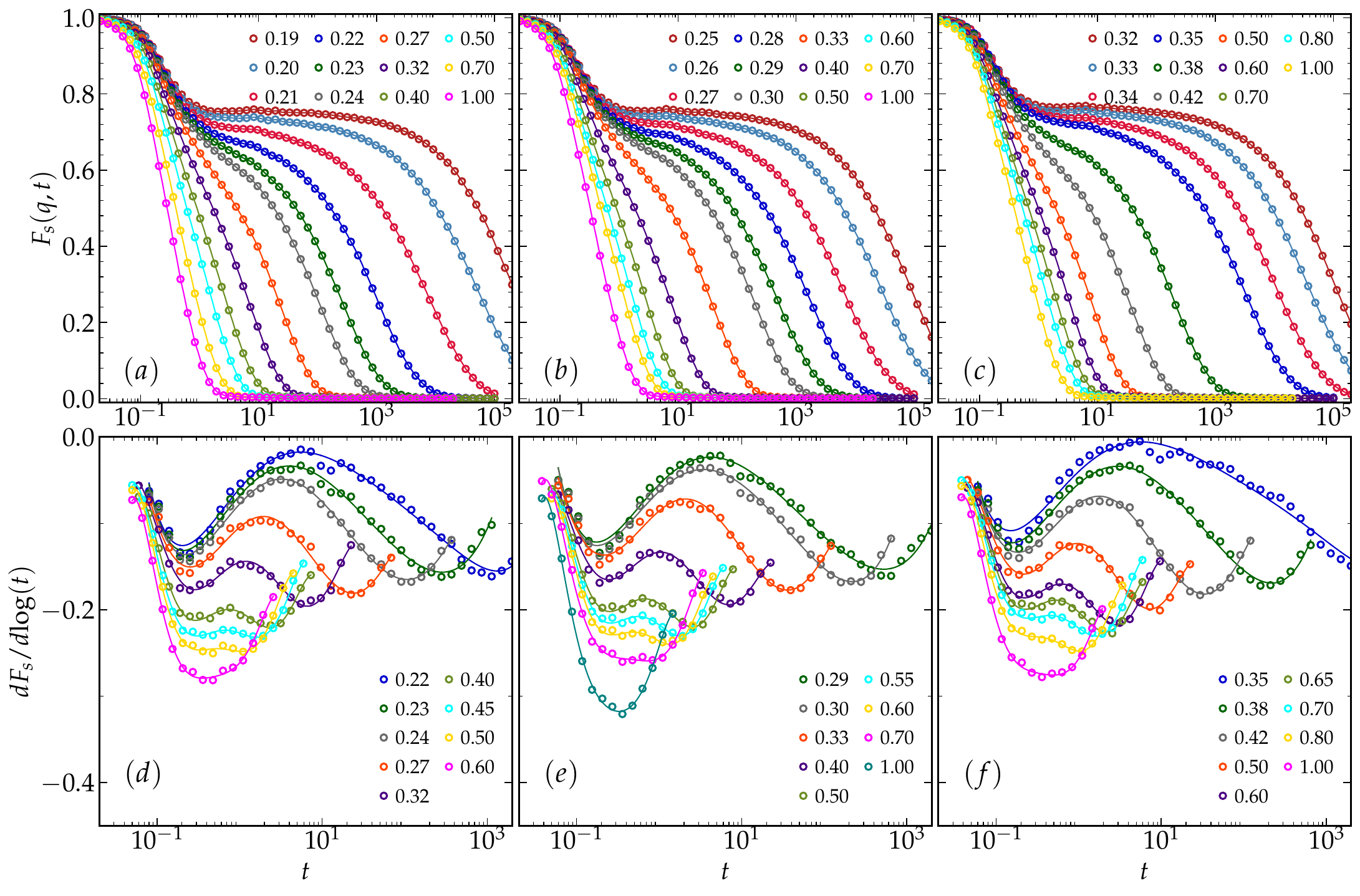}
\caption{Self-intermediate scattering function $F_s(q,t)$ at $q=6.5$ plotted as a function of time $t$ for the indicated temperatures. The upper panels (a-c) correspond to densities $\rho=0.75$, $0.80$, and $0.85$, respectively. The lower panels (d–f) show the corresponding time evolution of $tF’_s(q,t)$. Solid lines represent sixth-order polynomial fits to the derivative data.}
\label{fig:fsqt}
\end{figure*}

Inherent structures, corresponding to basins of potential-energy minima, are used to estimate the vibrational entropy of the system. Such structures are prepared by minimizing the potential energy of the corresponding equilibrated supercooled configurations via the conjugate-gradient method under constant volume. To ensure convergence to a local minimum, a tolerance threshold of $10^{-16}$ is applied to the energy. During the preparation of each inherent structure, it is verified that the potential-energy threshold is consistently achieved. Each configuration $\mathbf{r}^N$ of a supercooled liquid can be mapped to a corresponding local minimum, or inherent structure $\mathbf{r}^N_{\mathrm{IS}}$. By expanding the potential energy to second order around these minima, and within the harmonic approximation,
\begin{equation}
U_p(\mathbf{r}^N) \simeq U_p(\mathbf{r}^N_{\mathrm{IS}}) + \frac{1}{2} \sum_{i,j} \left. \frac{\partial^2 U_p}{\partial {\bf r}_i \partial {\bf r}_j} \right|_{\mathrm{IS}} \Delta {\bf r}_i \Delta {\bf r}_j.
\end{equation}
Here, \(U_p(\mathbf{r}_{\mathrm{IS}}^{N})\) is the $N$-particle inherent structure energy, and \(\Delta \mathbf{r}_{i}=\mathbf{r}_{i}-\mathbf{r}_{i,\mathrm{IS}}\) is the displacement of particle \(i\) from that minimum. The matrix $\frac{\partial^2 U_p}{\partial \mathbf{r}_i \partial \mathbf{r}_j} |_{\mathrm{IS}}$ is the Hessian, or curvature matrix, of the potential energy evaluated at the inherent-structure minimum. After diagonalizing the Hessian matrix yields $3N-3$ non-zero eigen-values $\lambda_k$, which define the normal-mode frequencies $\{\omega_k\} = \sqrt{\lambda_k}$ \cite{das2022crossover}. We exclude the three zero eigenvalues corresponding to global translations. The vibrational entropy is then calculated by averaging over equilibrium-sampled inherent structures \cite{das2022crossover},
\begin{equation}
S_{\mathrm{vib}}(T) = \left\langle \sum_{k=1}^{3N-3} \left[ 1 - \ln (\beta \hbar \omega_k) \right] \right\rangle_{\mathrm{IS}}.
\end{equation}
This expression assumes the high-temperature classical limit for the vibrational modes, a standard convention for molecular dynamics studies of supercooled liquids.

Fig.~\ref{Sc}(a) illustrates the total and vibrational entropies as a function of temperature across all densities, calculated using the aforementioned methodology. Although both quantities are monotonically increasing functions of temperature, the vibrational entropy exhibits a comparatively weaker temperature dependence. In Fig.~\ref{Sc}(b) we plotted the configurational entropy, calculated via Eq.~\eqref{eq:conf_ent}.

We also evaluate the average inherent structure energy per-particle, $ e_{\rm IS}  = \langle U_p(\mathbf{r}^N_{\mathrm{IS}}) \rangle/N$, where $\langle \cdot \rangle$ denotes average over independently prepared inherent structures sampled at different temperature for each density. In Fig.~\ref{IS}, we plot the normalized value of the inherent energy dinfed as $ \bar{e}_{\rm IS}= e_{\rm IS}/ e_{\rm IS}(T=1)$, where $e_{\rm IS}(T=1) = 1.748,\ 2.288,$ and $2.94$ for densities $\rho=0.75,\ 0.80,$ and $0.85$, respectively. It may be noted that below a temperature $T_{\rm IS}$ shown in the figure by the arrow, $\bar{e}_{\rm IS}$ deviates from its high temperature value of one \cite{banerjee2017determination,e_IS_2002}. The temperature $T_{\rm IS}$ can be regarded as a crossover temperature as it separates high temperatures from that of the low temperatures. We list the value of $T_{\rm IS}$ in Table~\ref{table} for the three densities.

\subsection{Structural relaxation time}\label{Sec_relaxation}
We determine the structural relaxation time, $\tau_{\alpha}$, using the self-intermediate scattering function and the self-overlap correlation function \cite{binder2011glassy}. The self-intermediate scattering function, $F_s(q,t)$, is computed from single-particle trajectories as
\begin{equation}\label{Fsqt}
F_s(q, t) = \frac{1}{N} \left\langle \sum_{i=1}^{N} \exp\left(-i\vec{q} \cdot \left[\vec{r}_i(t+t_0) - \vec{r}_i(t_0)\right]\right) \right\rangle,
\end{equation}
where $\vec{r}_i(t)$ denotes the position of particle $i$ at time $t$, and $t_0$ is a reference time origin. The angular brackets $\langle \cdot \rangle$ indicate averaging over independent trajectories and multiple time origins. To probe structural relaxation, the wavevector magnitude $q$ is chosen close to the first peak of the static structure factor. The $\alpha$-relaxation time, $\tau_{\alpha}$, is defined as the time at which the correlation decays to $1/e$ of its initial value, i.e., $F_s(q, \tau_{\alpha}) = 1/e$.

Figs.~\ref{fig:fsqt} (a-c) show the time dependence of $F_s(q,t)$ at different temperatures for the three densities studied. At high temperatures, the decay is characterized by a single step, whereas upon cooling it becomes progressively slower and develops a clear two-step relaxation pattern. Although not shown here, the intermediate- to long-time decay of $F_s(q,t)$ can be well described by stretched exponential fits \cite{vaibhav2022finite,trachenko2021slow,Pastore_2026}. The resulting $\alpha$-relaxation times obtained from $F_s(q,t)$ are presented in Fig.~\ref{relaxation} for all three densities.

The self-overlap correlation function \cite{karmakar2009growing} is defined as
\begin{equation}\label{overlap1}
q_s(t) = \frac{1}{N} \left\langle \sum_{i=1}^{N} \delta\big(\vec{r}_i(t+t_0) - \vec{r}_i(t_0)\big) \right\rangle,
\end{equation}
where the Dirac delta function is approximated by a window function $w(x)$, defined as $w(x)=1$ for $x \le a$ and $w(x)=0$ otherwise. This leads to
\begin{equation}
q_s(t) = \frac{1}{N} \left\langle \sum_{i=1}^{N} w\big(|\vec{r}_i(t+t_0) - \vec{r}_i(t_0)|\big) \right\rangle.
\end{equation}
A typical choice for the cutoff is $a = 0.3$. Analogous to $F_s(q,t)$, the $\alpha$-relaxation time is defined as the time at which $q_s(t)$ decays to $1/e$. We have verified that the relaxation times obtained from both $F_s(q,t)$ and $q_s(t)$ are consistent across all temperatures and densities reported in this study.

\begin{figure}[b]
\includegraphics[width=0.45\textwidth,clip]{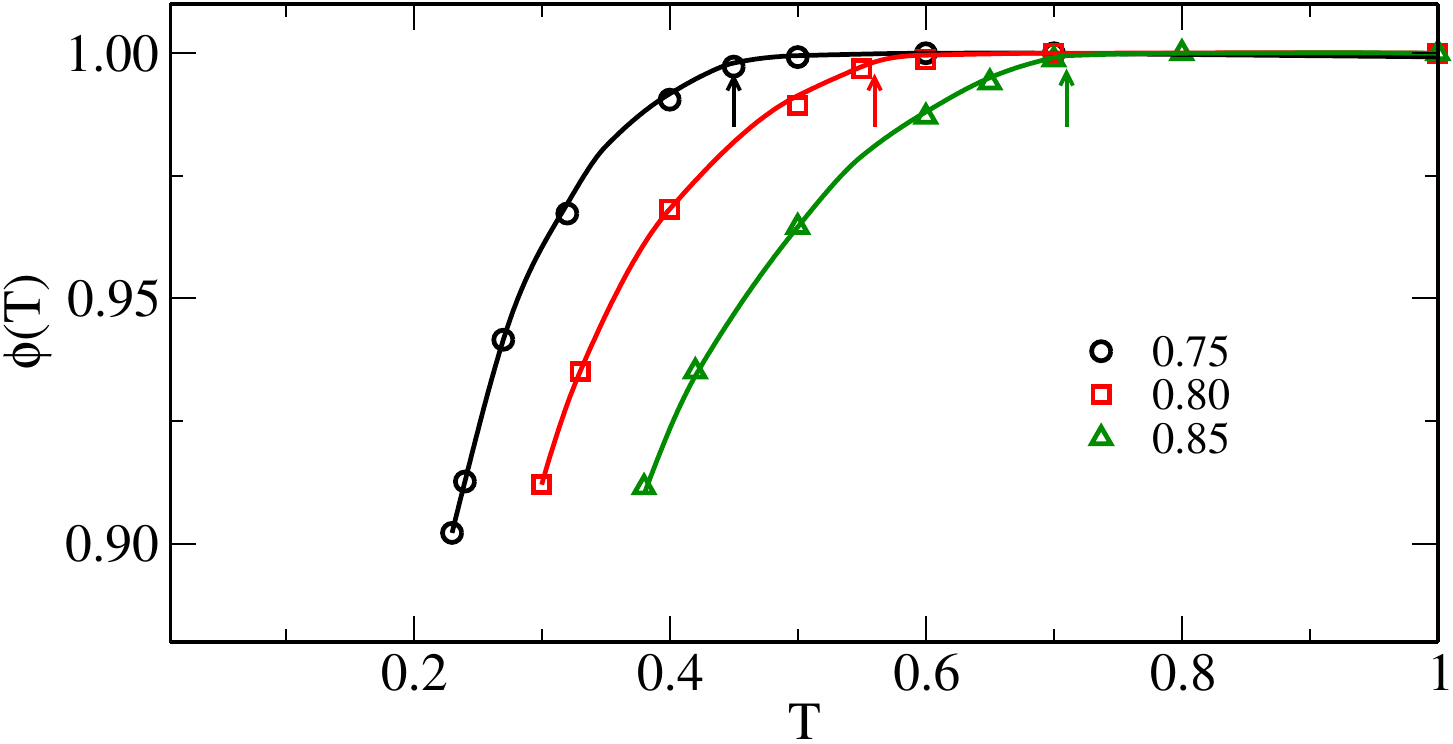}
\caption{We plotvi the parameter $\phi(T)$ as a function of temperature for the three densities. Parameter $\phi(T)=1-h(T)$, is obtained from the two-minima structure of $tF'_s(q,t)$, in which the two minima are separated by a barrier $h(T)$. Symbols represent the calculated values, while the solid lines represent the corresponding cubic fit. The arrow indicates the crossover temperature $T_{h}$ for different densities, which are listed in Table \ref{table}.}
\label{phi_T}
\end{figure}

We calculate the logarithmic derivative of $F_s(q,t)$ ~\cite{coslovich2025freezing}, 
\begin{equation}
\frac{dF_s(q, t)}{d \log t} = t F'_s(q, t).
\end{equation}
The derivative $t F'_s(q,t)$ has been calculated using the central-difference method and subsequently fitted with a sixth-order polynomial in $\log t$, restricting the fit to the interval $0.05 < F_s(q,t) < 0.98$. In Fig.~\ref{fig:fsqt}, we plot the time evolution of $tF'_s(q,t)$ at different temperatures for densities $\rho=0.75$, $0.80$, and $0.85$, as shown in panels (d), (e), and (f), respectively. The symbols represent the calculated values, while the solid lines represent the sixth-order polynomial fits to the data. Below a temperature denoted as $T_h$, a two-minima structure of $t F'_s(q, t)$ separated by a barrier appears. We define a quantity $\phi(T)=1-h(T)$, where $h(T)$ is the height of the barrier, and plot it in Fig.~\ref{phi_T} as a function of $T$ for the three densities. An arrow in the figure marks the point at which $\phi(T)$ departs from its high-temperature value of $1$ and starts decreasing upon lowering the temperature. For each density the marked point defines a temperature $T_h$, which separates the way the quantity $F_s(q,t)$ depends on time $t$ in the high-temperature region from that in the low-temperature region. Symbols represent the calculated values, while the solid lines represent the corresponding cubic fit. The arrow indicates the crossover temperature $T_{h}$ for different densities, which are listed in Table \ref{table}.

\begin{table}[b]
\caption{\label{table} Characteristic temperatures representing the crossover temperature obtained from different approaches for three densities.}

\begin{ruledtabular}
\begin{tabular}{cccc}
Density ($\rho$) & $T_{\rm IS}$ & $T_a$ & $T_{h}$ \\
\hline
$0.75$ & 0.39  &0.39  &0.45  \\
$0.80$ & 0.49  &0.50  &0.56  \\
$0.85$ & 0.62  &0.64  &0.71  \\
\end{tabular}
\end{ruledtabular}
\end{table}

\section{Theory and results} \label{theory}
\begin{figure}[t]
\includegraphics[width=0.48\textwidth,clip]{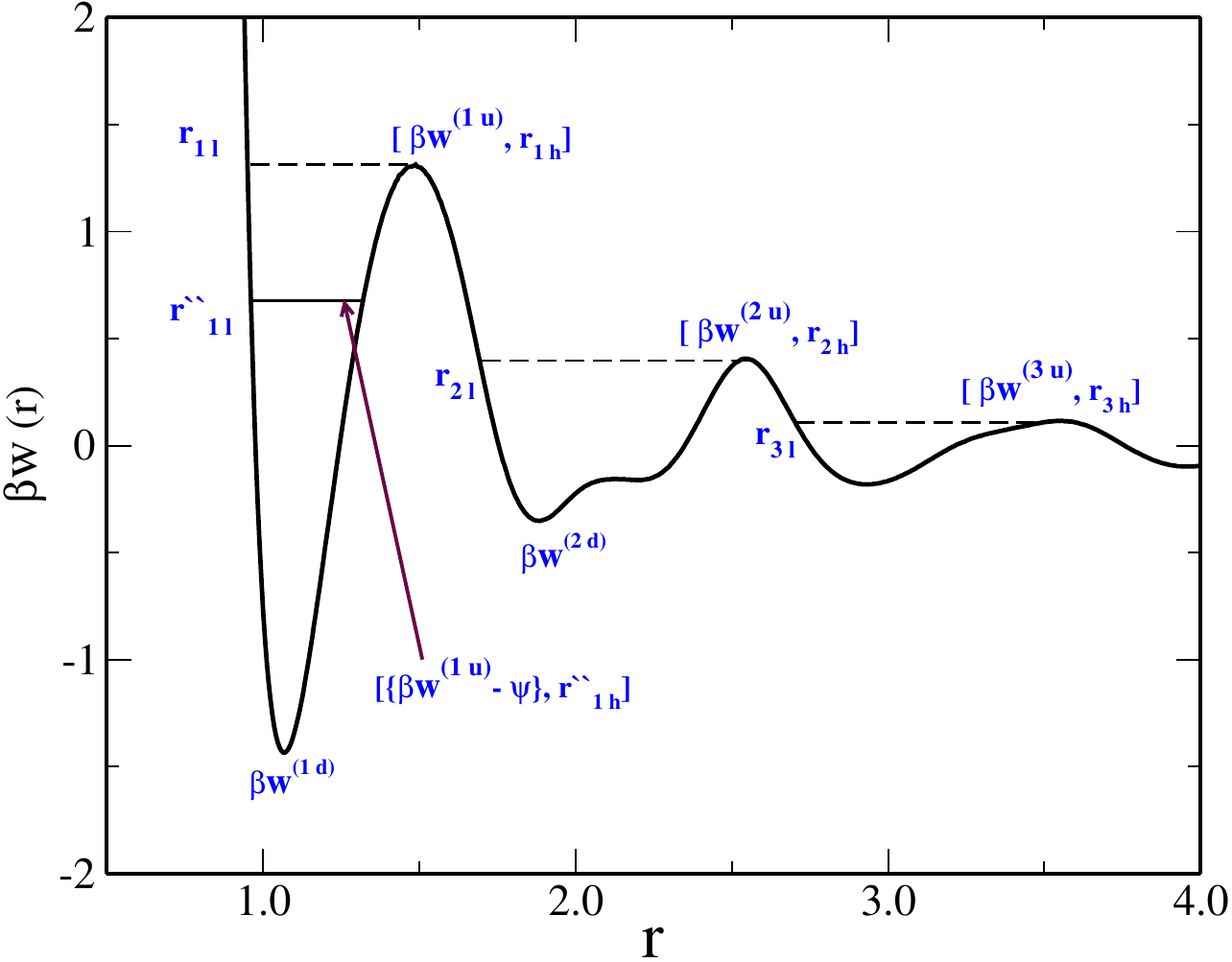}
\caption{The reduced effective potential $\beta w(r)$ between a pair of particles of species $\alpha$ and $\gamma$ separated by distance $r$ in a system of IPL potential at a density $\rho=0.80$ and temperature $T=0.26$. $\beta w^{(iu)}, r_{ih}$ are, respectively, value and location of $i$th maximum and $r_{il}$ is the corresponding location on the left hand side of the shell where $\beta w^{(i)}(r)=\beta w^{(iu)}$ (shown by dashed line). The location $r''_{il}$ and $r''_{ih}$ are values of $r$ on the left and the right hand side of the shell where $\beta w^{(i)}(r)=[\beta w^{(iu)}-\psi]$ (shown by full line). $\beta w^{(id)}$ is the depth of the $i$th shell.}
\label{fig-1}
\end{figure}
\subsection{Outline of the theory\label{theoryA}}

\begin{figure}[t]
    \includegraphics[width=0.45\textwidth,clip]{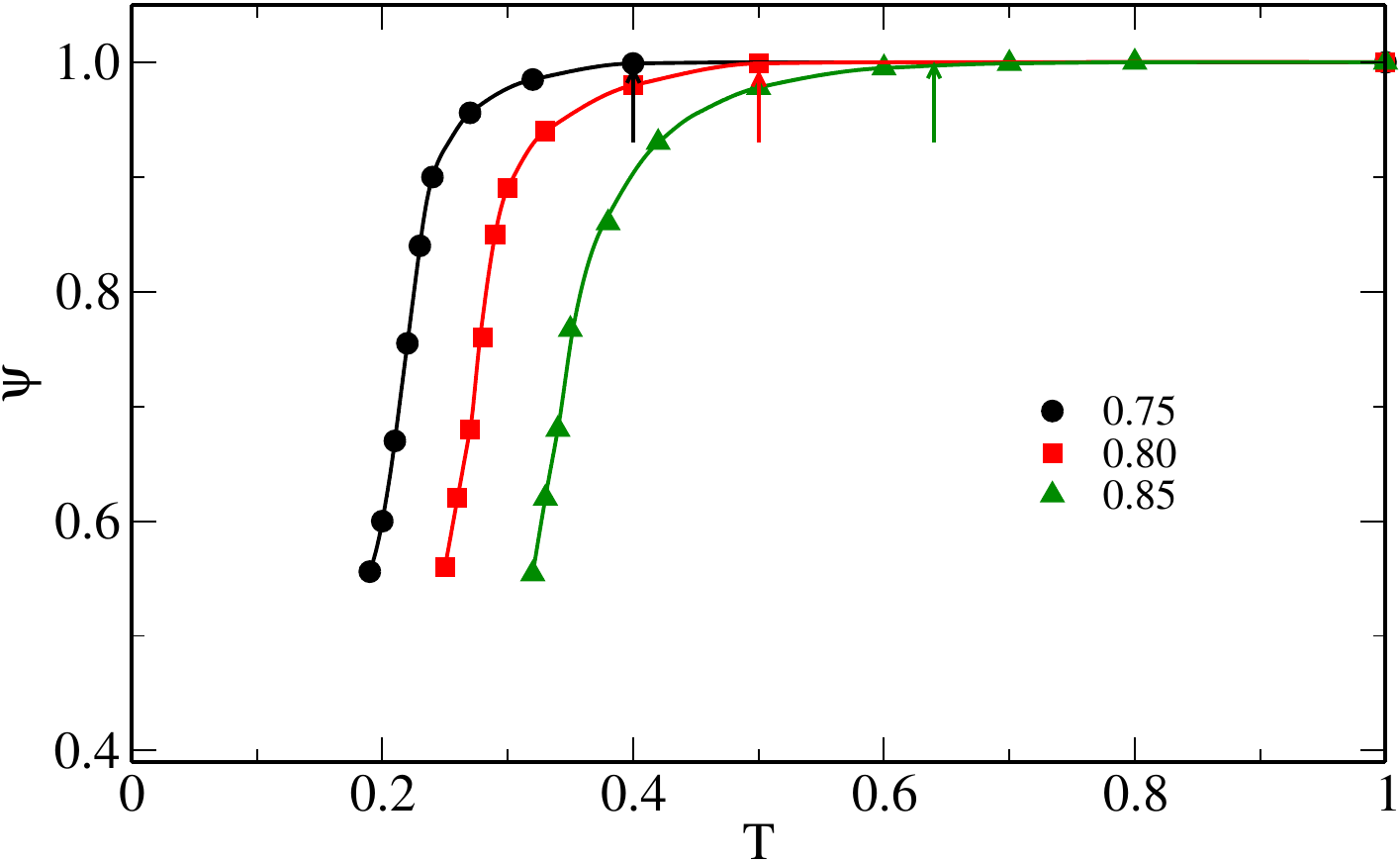}
    \caption{Dependence of $\psi$ on temperature $T$ for the densities $\rho = 0.75, 0.80$ and $0.85$. Symbols represent the calculated values, while the solid lines represent the corresponding least square fit. The arrow indicates the crossover temperature $T_a$ for different densities, which are listed in Table \ref{table}.}
    \label{fig-2}
\end{figure}

Our theory~\cite{PhysRevE.103.032611, PhysRevE.103.052105, PhysRevE.99.030101, PhysRevE.107.014119}  provides a method to identify and calculate the numbers of dynamically free, metastable, and stable neighbors of a tagged (central) particle in a fluid. To achieve this, the definition of $g_{\alpha\gamma}(r)$ is modified to include the momentum distribution~\cite{PhysRevE.99.030101,PhysRevE.103.032611};
\begin{equation}
 g_{\alpha\gamma}(r)=\left(\frac{\beta}{2\pi\mu}\right)^{\frac{3}{2}}\int \mathrm{d}{\bf p} \  
\mathrm{e}^{-\beta (\frac{p^2}{2\mu} + w_{\alpha\gamma}(r))}, 
\end{equation}
where ${\bf p}$  is the relative momentum of a particle of reduced mass $\mu=m/2$,  $\beta=(k_{B}T)^{-1}$ is the inverse temperature measured in units of Boltzmann constant $k_{B}$. A particle in the shell interacts with the central particle with an effective potential (potential of mean force \cite{Hansen}) $w(r)=-k_{B}T\ln g(r)$, which is the sum of the (bare) potential energy $u(r)$ and the system-induced potential energy of interaction between a pair of particles of species $\alpha$ and $\gamma$ separated by distance $r$.

In Fig.~\ref{fig-1}, we plot $\beta w_{\alpha\gamma}(r)$ between a pair of particles of species $\alpha$ and $\gamma$ separated by distance $r$ (expressed in unit of $\sigma_{AA}$), for a fluid interacting via IPL potential given by Eq.~\eqref{potential-n} at a number density $\rho=0.80$ and temperature $T=0.26$. We identify several potential wells separated by barriers. A region between two barriers (maxima), leveled as $i-1$ and $i$, where $(i\geq 1)$, is dubbed as $i$th shell (cage) and the minimum of the shell is denoted as $\beta w_{\alpha\gamma}^{(id)}$. The value of the $i^{th}$ barrier is denoted as $\beta w_{\alpha\gamma}^{(iu)}$ and its location is denoted by $r_{ih}$.

The number of particles trapped in the $i$th shell is found
\begin{eqnarray}\nonumber
g_{\alpha\gamma}^{(ib)}(r)  &=& 4\pi(\frac{\beta}{2\pi\mu})^{3/2} \mathrm{e}^{-\beta w_{\alpha\gamma}^{(i)}(r)} 
\int_{0}^{\sqrt{2\mu[w_{\alpha\gamma}^{(iu)}-w_{\alpha\gamma}^{(i)}(r)]}}      \\
&&  \times \mathrm{e}^{-\beta p^2/2\mu} p^2 \mathrm{d}p ,
\end{eqnarray}
where $w_{\alpha\gamma}^{(i)}(r)$ is the effective potential in the range of $r_{il}\leq r \leq r_{ih}$ of $i$th shell, $r_{il}$ is value of $r$ where $w_{\alpha\gamma}^{(i)}(r)= w_{\alpha\gamma}^{(iu)}$ on the left-hand side of the shell (Fig.~\ref{fig-1}). The number of particles bonded with the central particle of species $\alpha$ is
\begin{equation}\label{nb}
n_{\alpha}^{(b)} = 4\pi \sum_{i} \sum_{\gamma}\rho_{\gamma}\int_{r_{il}}^{r_{ih}} 
g_{\alpha\gamma}^{(ib)}(r)~r^2 \mathrm{d}r  ,
\end{equation}
where summations are over all shells and over all species, and $\rho_{\gamma}$ is number density of species $\gamma$. This number $n_{\alpha}^{(b)}$ increases rapidly on lowering the temperature and increasing the density due to increase in the number of shells surrounding the central particle and increase in values of maximum and minimum of each shell. The size of the cluster defines one of the length scale of the glassy dynamics.

\begin{figure}[t]
\includegraphics[width=0.43\textwidth,clip]{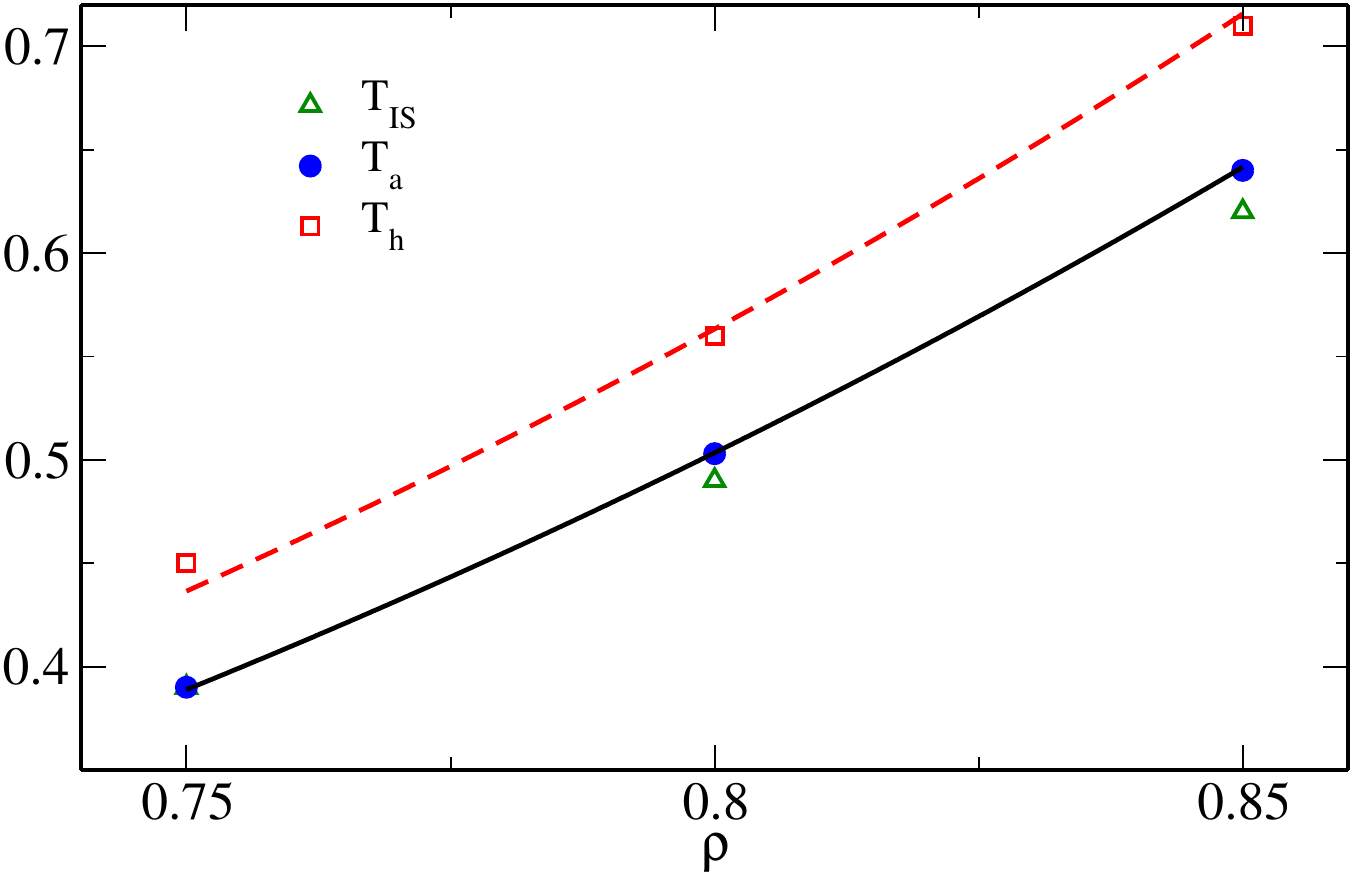}
\caption{Density dependence of the crossover temperatures $T_{IS}$, $T_a$, and $T_h$. Symbols represent the calculated values given in Table~\ref{table}, while the lines represent values found from relations $T_a = 1.23\rho^4$ (solid line) and $T_h = 1.37 \rho^4$ (dashed line).}
\label{Ta_scaling}
\end{figure}

\begin{figure*}[]
    \includegraphics[width=0.75\textwidth,clip]{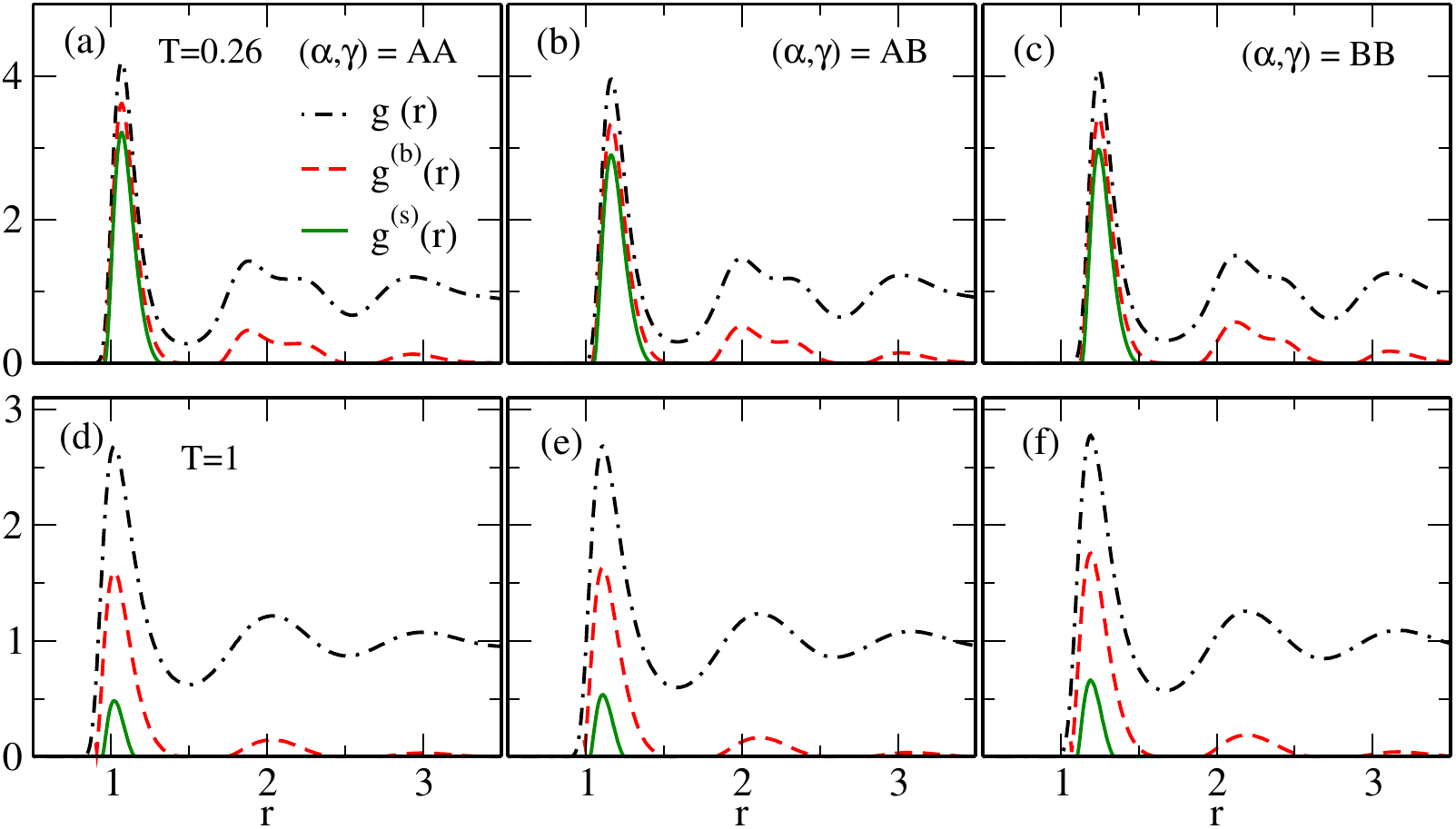}
    \caption{Comparison of $g_{\alpha \gamma}(r)$ (dot–dashed line), $g_{\alpha \gamma}^{(b)}(r)$ (dashed line), and $g_{\alpha \gamma}^{(s)}(r)$ (solid line). Panels (a–c) correspond to $T = 0.26$, while panels (d–f) correspond to  $T = 1.0$, all at density $\rho = 0.80$. The first, second, and third columns show the $A$–$A$, $A$–$B$, and $B$–$B$ correlations, respectively.}
    \label{fig-2a}
\end{figure*}

However, thermal fluctuations embedded in the system (bath) drive those particles whose energies are closer to the barrier height out of the shell. These particles are referred to as $m$ (metastable) particles. The bath fluctuations consist of two competing contributions: one is entropy-driven, which makes particle escape, and the other is energy-driven, which opposes the escape and diminishes the effect of entropy-driven fluctuations. In the theory, a temperature-dependent function $\psi(T)$ is introduced as a measure of the effect created on the cluster of bonded particles. Because of these fluctuations, all those particles of $i$th shell whose energies lie between  $\beta w_{\alpha\gamma}^{(i u)}-\psi$ and  $\beta w_{\alpha\gamma}^{(i u)}$ escape the shell are $m$-particles. On the other hand, all those particles whose energies are lower than $[\beta w_{\alpha\gamma}^{(i u)}-\psi]$  remain trapped in the shell, are $s$-particles. The number of $s$-particles at a given temperature and density is found from a part of $g_{\alpha\gamma}(r)$ defined as
\begin{eqnarray}\nonumber
g_{\alpha\gamma}^{(is)}(r)  &=& 4\pi(\frac{\beta}{2\pi\mu})^{3/2} \mathrm{e}^{-\beta w_{\alpha\gamma}^{(i)}(r)} 
\int_{0}^{\sqrt{2\mu[w_{\alpha\gamma}^{(iu)}-\psi k_{B} T-w_{\alpha\gamma}^{(i)}(r)]}}      \\
&&  \ \times \mathrm{e}^{-\beta p^2/2\mu} p^2 \mathrm{d}p ,
\label{grs}
\end{eqnarray}
where $w_{\alpha\gamma}^{(i)}(r)$ is in the range $r''_{il}\leq r \leq r''_{ih}$. Here $r''_{il}$ and $r''_{ih}$ are, respectively, value of $r$ on the left and the right hand side of the shell where $\beta w_{\alpha\gamma}^{(i)}(r)= \beta w_{\alpha\gamma}^{(iu)}-\psi$. The number of $s$-particles around a $\alpha$ particle is 
\begin{equation}
n_{\alpha}^{(s)} = 4\pi \sum_{i} \sum_{\gamma}\rho_{\gamma}\int_{r''_{il}}^{r''_{ih}} 
g_{\alpha\gamma}^{(is)}(r) r^2 \mathrm{d}r  .
\label{ns}
\end{equation}

The averaged number of $s$-particle bonded with a central particle in a binary mixture is
\begin{equation}
n^{(s)} = x_{A} n_{A}^{(s)} + x_{B} n_{B}^{(s)}  ,
\label{nst}
\end{equation}
where $x_{\alpha}$ is the concentration of species $\alpha$. 

The cluster of $n^{(s)}$ particles is identified as CRC. It may be noted that the particles of CRC are distributed in shells around the central particle where they share the region with metastable particles. The CRC differs from the Adam and Gibbs \cite{Adam} ``cooperatively rearranging region'' which is taken to be a compact structure \cite{Adam, Bouchaud}. The CRC is embedded at the centre of a much larger cluster of $m$-particles. Since $m$-particles are loosely bonded, they move individually or in a small groups on timescale much smaller than $\tau_{\alpha}$ without affecting the structure of CRC. But when CRC reorganizes at timescales of $\tau_{\alpha}$ it triggers the reorganization of all particles of the cluster turning into a large cluster of mobile particles. In simulations as well as in experiments, one may therefore observe relaxation governed by rapid sporadic events characterized by the emergence of a relatively large and compact cluster of mobile particles \cite{PhysRevLett.96.057801}. The glassy dynamics involves two length scales; one given by CRC and other by a cluster formed by all bonded of ($s$ plus $m$) particles.

\begin{figure}[b!]
    \includegraphics[width=0.4\textwidth,clip]{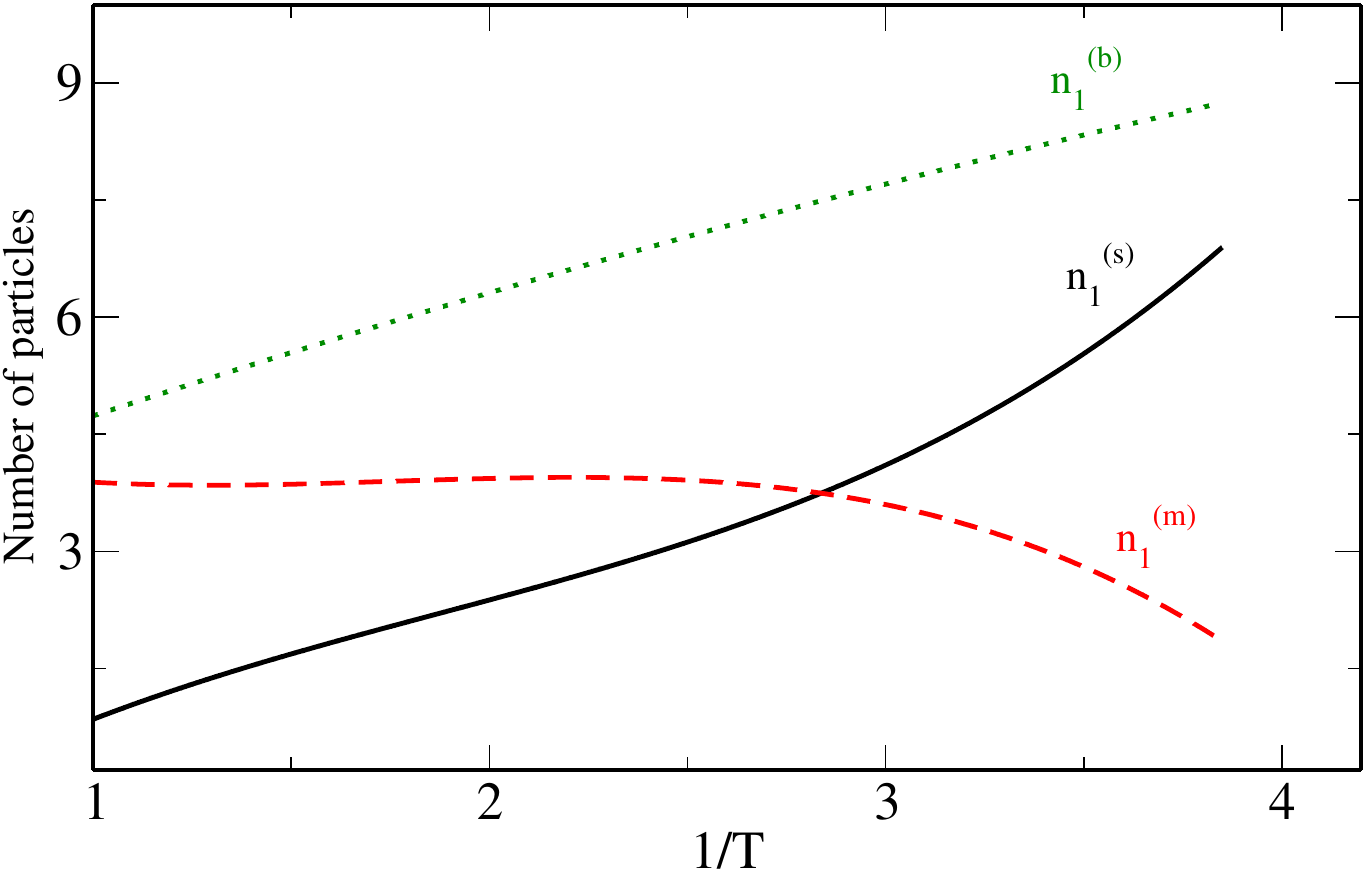}
    \caption{Number of bonded $n_{1}^{(b)}$ (dotted line), metastable $n_{1}^{(m)}$ (dashed line), and stable $n_{1}^{(s)}$ (solid line) particles in the first coordination shell as a function of inverse temperature $1/T$ at density $\rho=0.80$.}
    \label{particle}
\end{figure}

Number of particles in the cluster is related to the configurational entropy $S_{c}$ through the Adam and Gibbs \cite{Adam} relation,
\begin{equation}\label{K}
n^{(s)}(T)+1 = \dfrac{K}{S_{c}(T)} \ \ ,
\end{equation}
where $K$ is a temperature independent constant. A method for determining values of K and $\psi(T)$ using known value of $S_{c}$, and Eqs.~\eqref{ns}-\eqref{Es} is given in Ref.~ \cite{PhysRevE.103.032611, PhysRevE.103.052105}.

For an event of structural relaxation to take place the CRC has to reorganize irreversibly; the energy involved in this process is the effective activation energy $\beta E^{(s)}$ of relaxation. It is equal to the energy with which the central particle is bonded with the $s$-particles and is given as \cite{PhysRevE.103.032611, PhysRevE.103.052105}
\begin{eqnarray}\nonumber \label{Es}
\beta E^{(s)}(T,\rho) &=& 4\pi \sum_{i}\sum_{\gamma}x_{\gamma}\rho_{\gamma} \int_{r''_{il}}^{r''_{ih}} 
[\beta w_{\alpha\gamma}^{(iu)}-\psi(T)- \beta w_{\alpha\gamma}^{(i)}(r)] \\
&& \ \times g_{\alpha\gamma}^{(is)}(r) r^2 \mathrm{d}r ,
\end{eqnarray}
where energy is measured from the effective barrier $\beta w_{\alpha\gamma}^{(iu)}-\psi(T)$. The structural relaxation time $\tau_{\alpha}$ is obtained from the Arrhenius law,
\begin{equation}
\tau_{\alpha}(T,\rho) = \tau_{0} \exp{\left[\beta E^{(s)} (T,\rho)\right]}  ,
\label{tau_E}
\end{equation}
where $\tau_{0}$ is a microscopic time scale.

\begin{figure}[t!]
\includegraphics[width=0.4\textwidth,clip]{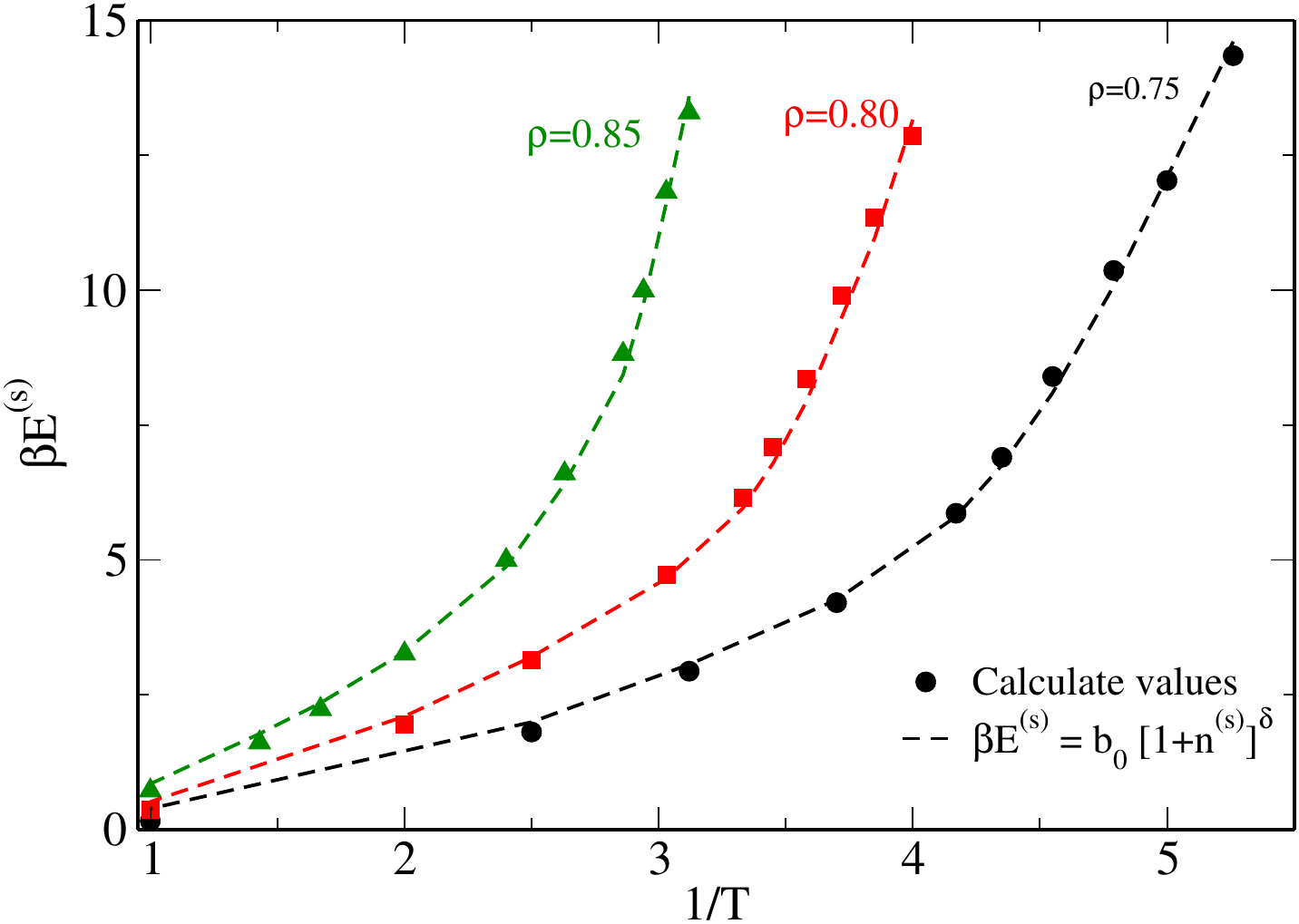}
\caption{Activation energy $\beta E^{(s)}$ of relaxation as a function of inverse temperature $1/T$ for densities $\rho = 0.75$, $0.80$, and $0.85$. Symbols denote the calculated values, while dashed lines show equation fits of the form $\beta E^{(s)} = b_{0} \,[1 + n^{(s)}]^{\delta}$, with $b_{0} = 0.18$ and $\delta = 2.0$, where $n^{(s)}$ is the number of bonds in a CRC.}
\label{energy}
\end{figure}

\subsection{Results and comparison with simulation}\label{results}

Calculated values of $\psi(T)$ is plotted in Fig.~\ref{fig-2} as a
function of $T$ for densities, $\rho = 0.75$, $0.80$, and $0.85$. The point marked by an arrow in Fig.~\ref{fig-2} separates the high-temperature dependence of $\psi(T)$ on $T$ from that of low temperatures. For $T>T_a$, $\psi(T)=1$, a value equal to that found in athermal glass formers, at $T=T_a$, $\psi(T)$ takes a turn and starts decreasing rather sharply, for $T<T_a$, as was found for the LJ \cite{PhysRevE.103.032611} and WCA \cite{PhysRevE.103.052105} glass formers. In Table~\ref{table}, we compare the values of $T_a$ with $T_h$ and $T_{\mathrm{IS}}$. As discussed above, they represent temperatures at which crossover from high-temperature behaviour to low-temperature behaviour takes place. While the values of $T_{IS}$ and $T_a$ are close, $T_h$ is about 12\% higher compare to $T_a$. This indicates that the dynamics is more sensitive to changes in the nature of fluctuations compared to thermodynamics. As pointed out above, the underlying cause for the crossover is dominance of energy driven fluctuations over entropy driven fluctuations. In Fig.~\ref{Ta_scaling}, we plot $T_{IS}$, $T_a$, and $T_h$ to show their density dependence. In the figure lines represent the values calculated from the relations $T_a = 1.23\rho^4$ (solid line) and $T_h = 1.37 \rho^4$ (dashed line).

In Fig.~\ref{fig-2a}, we compare $g_{\alpha \gamma}(r)$, $g_{\alpha \gamma}^{(b)}(r)$, and $g_{\alpha \gamma}^{(s)}(r)$ for $T=0.26$ and $T=1$ at $\rho=0.80$. While $g_{\alpha \gamma}^{(s)}(r)$ is short-ranged function limited to the first neighbor, $g_{\alpha \gamma}^{(b)}(r)$ range is as large as that of $g_{\alpha \gamma}(r)$. We also show the different components of the $g_{\alpha \gamma}(r)$ in the upper panel of the figure at low temperature $T = 0.26$ and in the lower panel at higher temperature $T = 1.0$. From the figure, it is clear that the length scale of $g_{\alpha \gamma}^{(b)}(r)$ and $g_{\alpha\gamma}^{(s)}(r)$ increases and becomes more pronounced at lower temperatures, which later is reflected in the number of particles.

\begin{figure}[t]
\includegraphics[width=0.4\textwidth,clip]{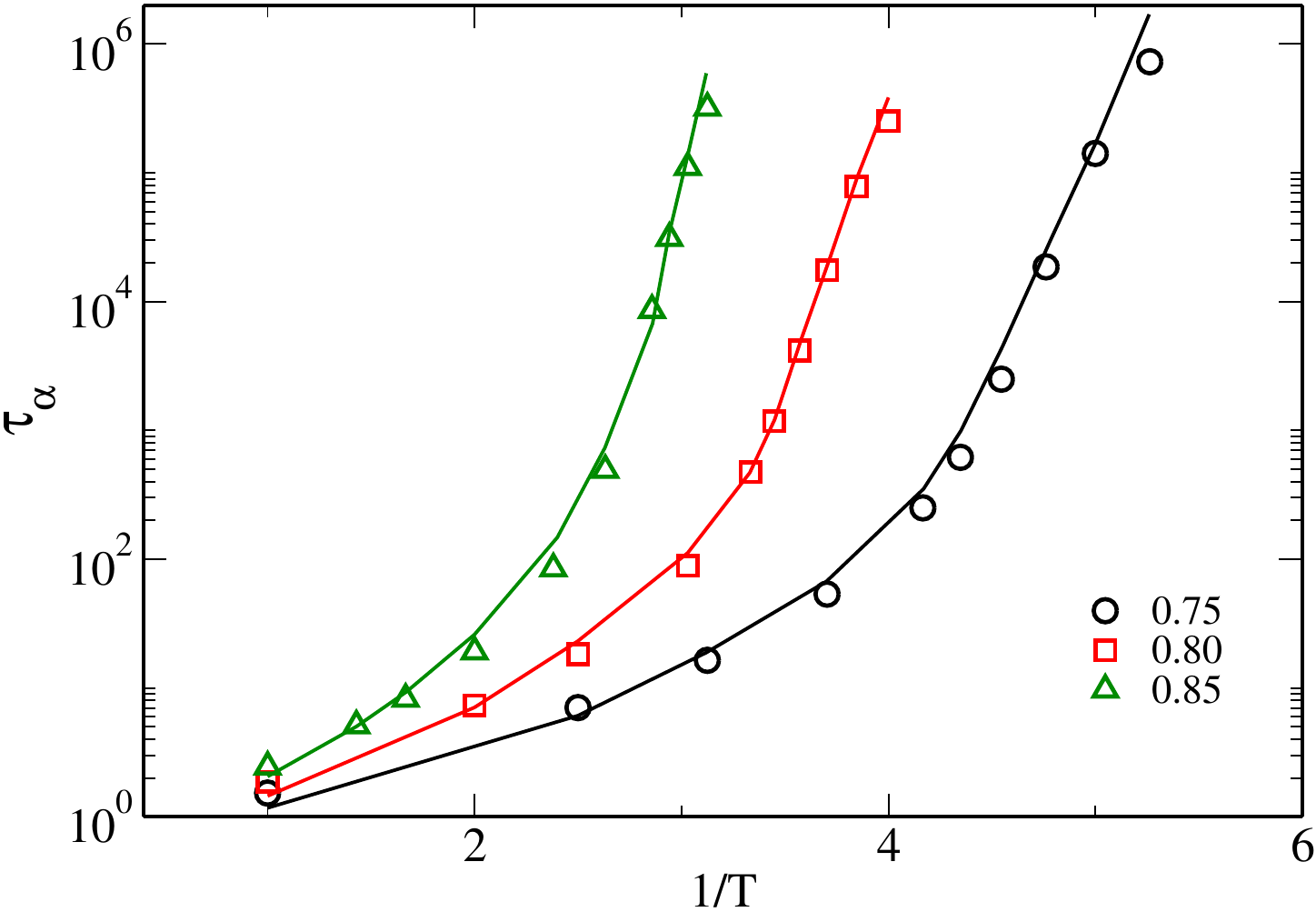}
\caption{Structural relaxation time $\tau_{\alpha}$ as a function of inverse temperature $1/T$ for densities $\rho = 0.75$, $0.80$, and $0.85$. Symbols represent the estimates obtained from the self-intermediate scattering function measured using MD simulations, while the solid lines indicate the theoretical predictions.}
\label{relaxation}
\end{figure}

\begin{figure*}[t]
    \centering

    \begin{minipage}[t]{0.33\textwidth}
        \centering
        \includegraphics[width=\linewidth]{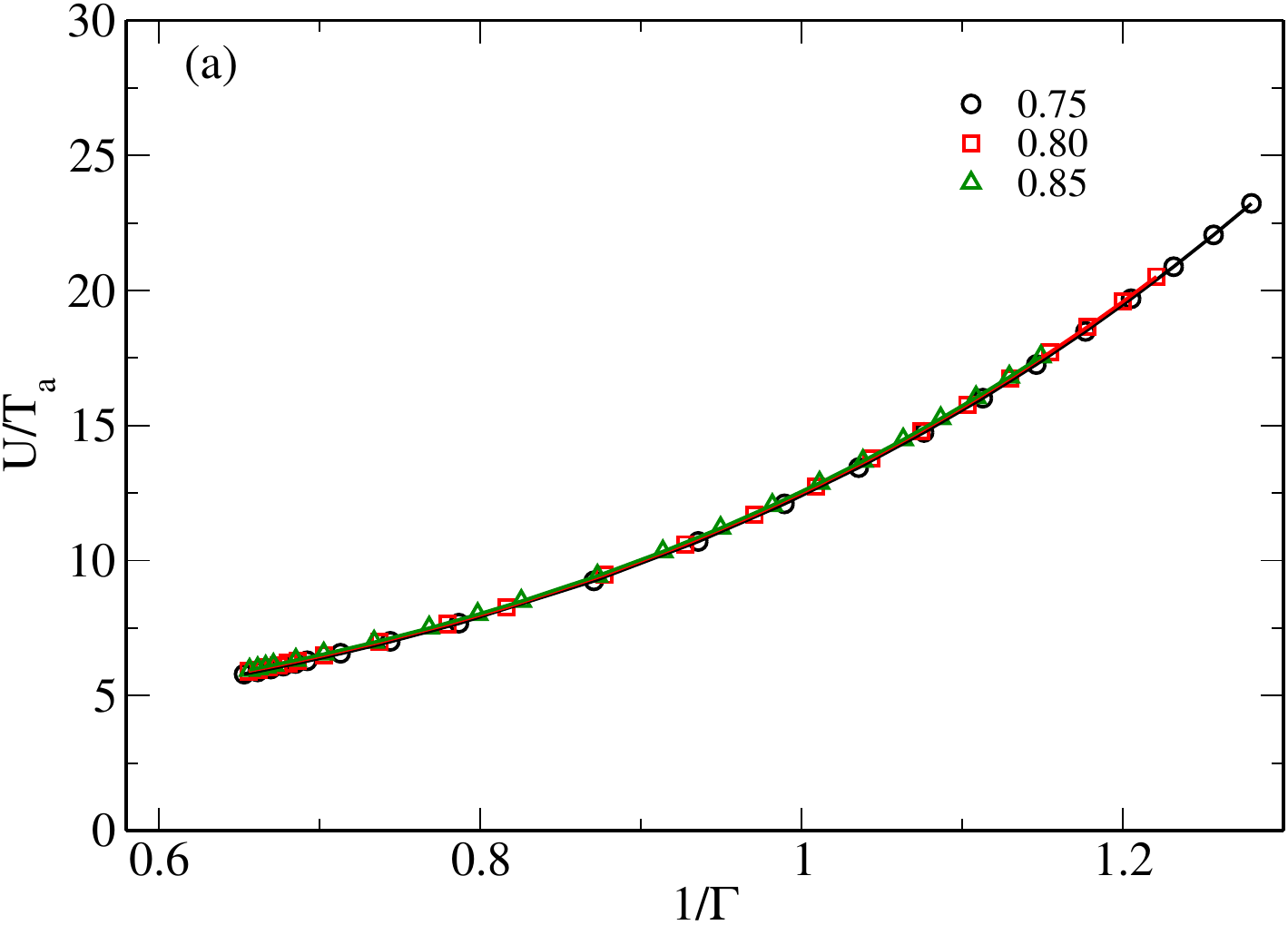}
    \end{minipage}
    \hfill
    \begin{minipage}[t]{0.33\textwidth}
        \centering
        \includegraphics[width=\linewidth]{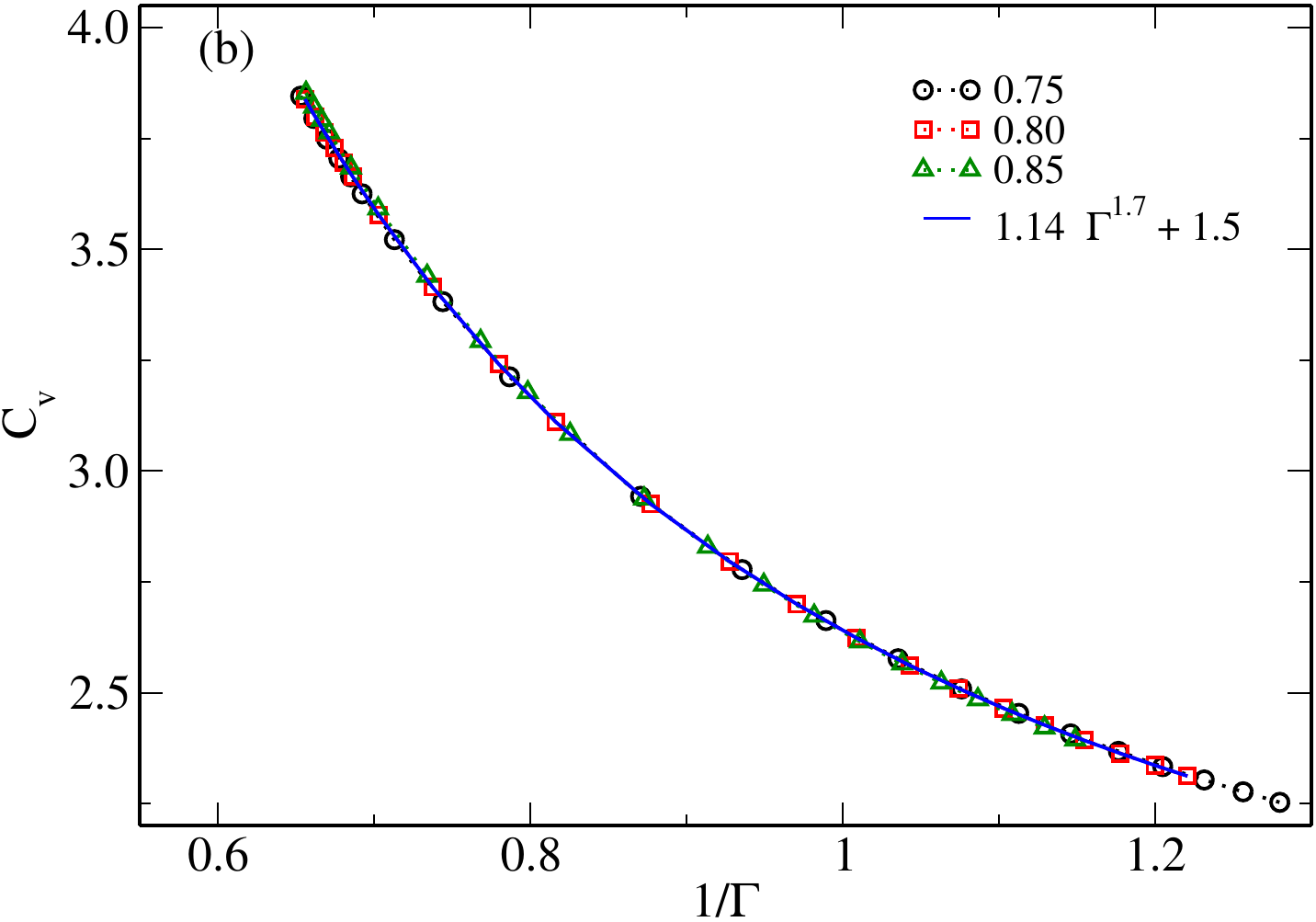}
    \end{minipage}
    \hfill
    \begin{minipage}[t]{0.33\textwidth}
        \centering
        \includegraphics[width=\linewidth]{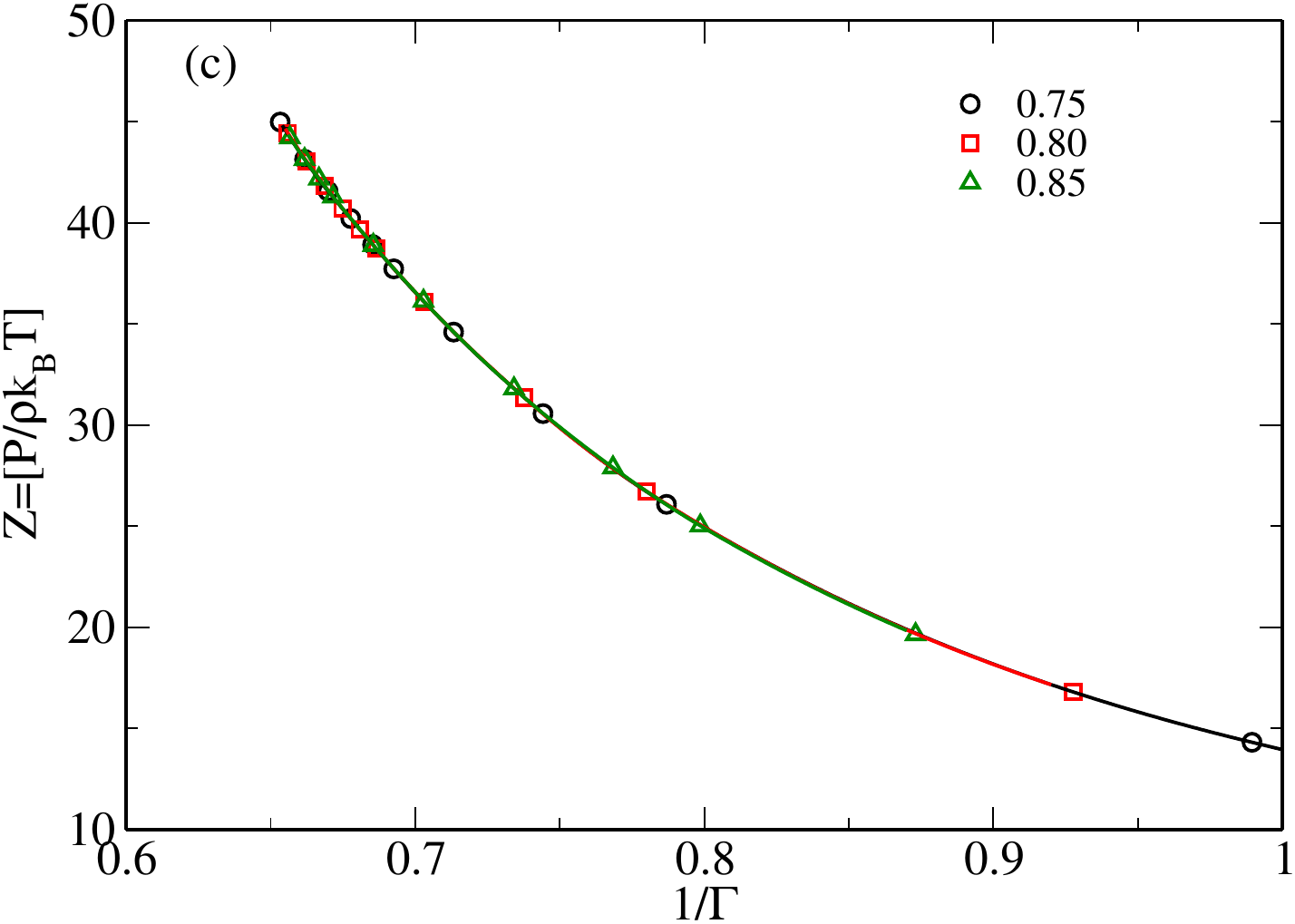}
    \end{minipage}
    \centering

    \begin{minipage}[t]{0.33\textwidth}
        \centering
        \includegraphics[width=\linewidth]{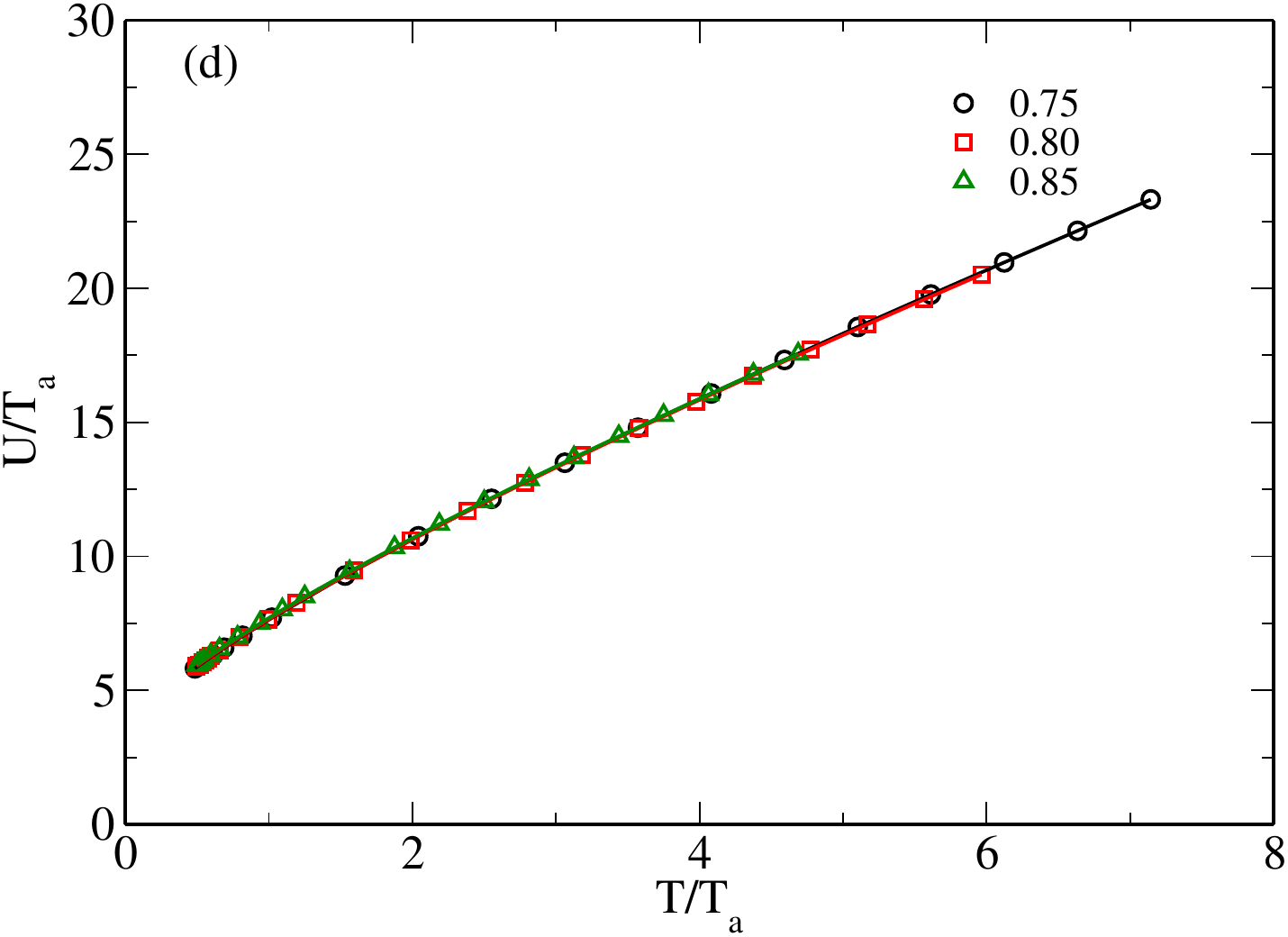}
    \end{minipage}
    \hfill
    \begin{minipage}[t]{0.33\textwidth}
        \centering
        \includegraphics[width=\linewidth]{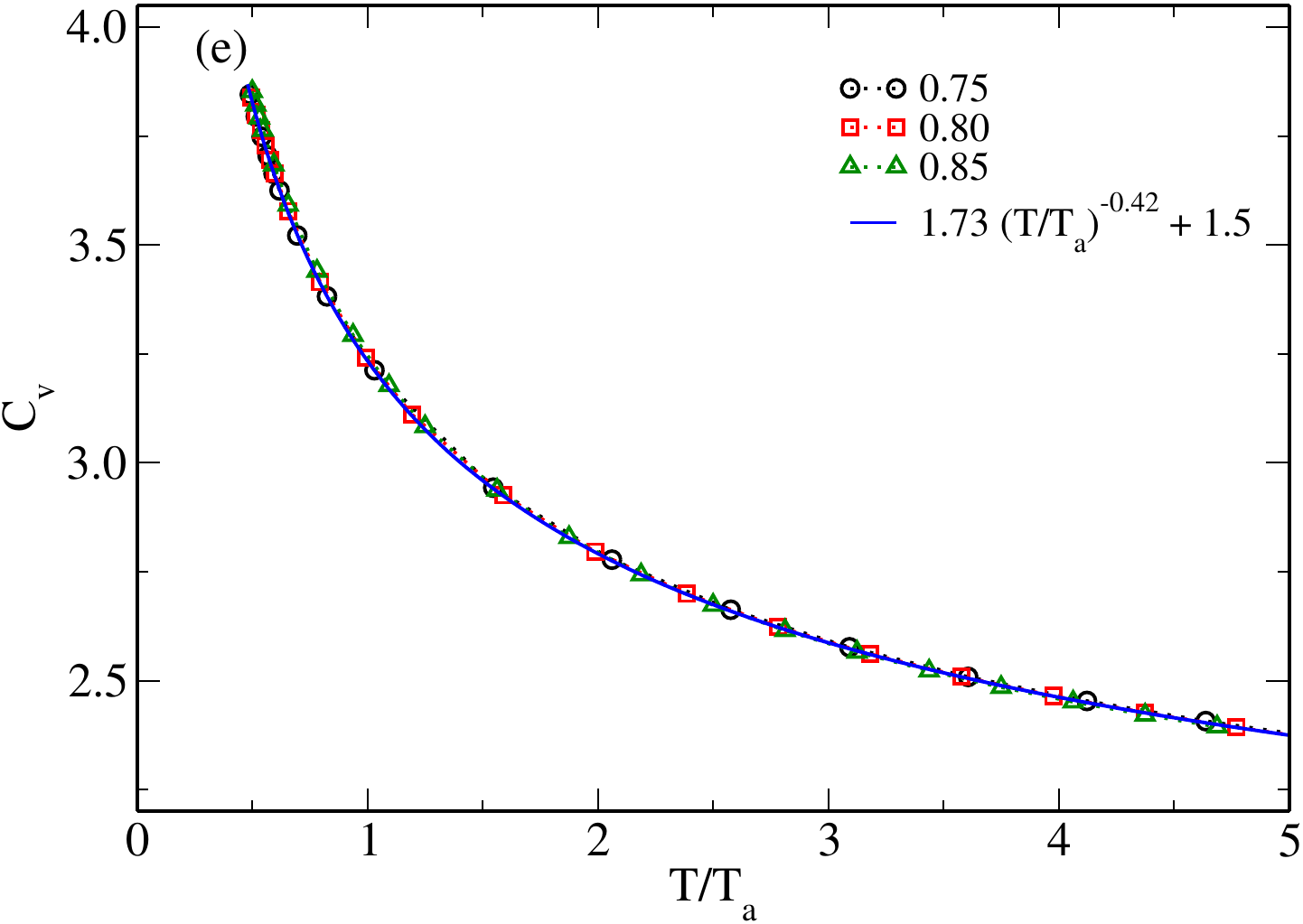}
    \end{minipage}
    \hfill
    \begin{minipage}[t]{0.33\textwidth}
        \centering
        \includegraphics[width=\linewidth]{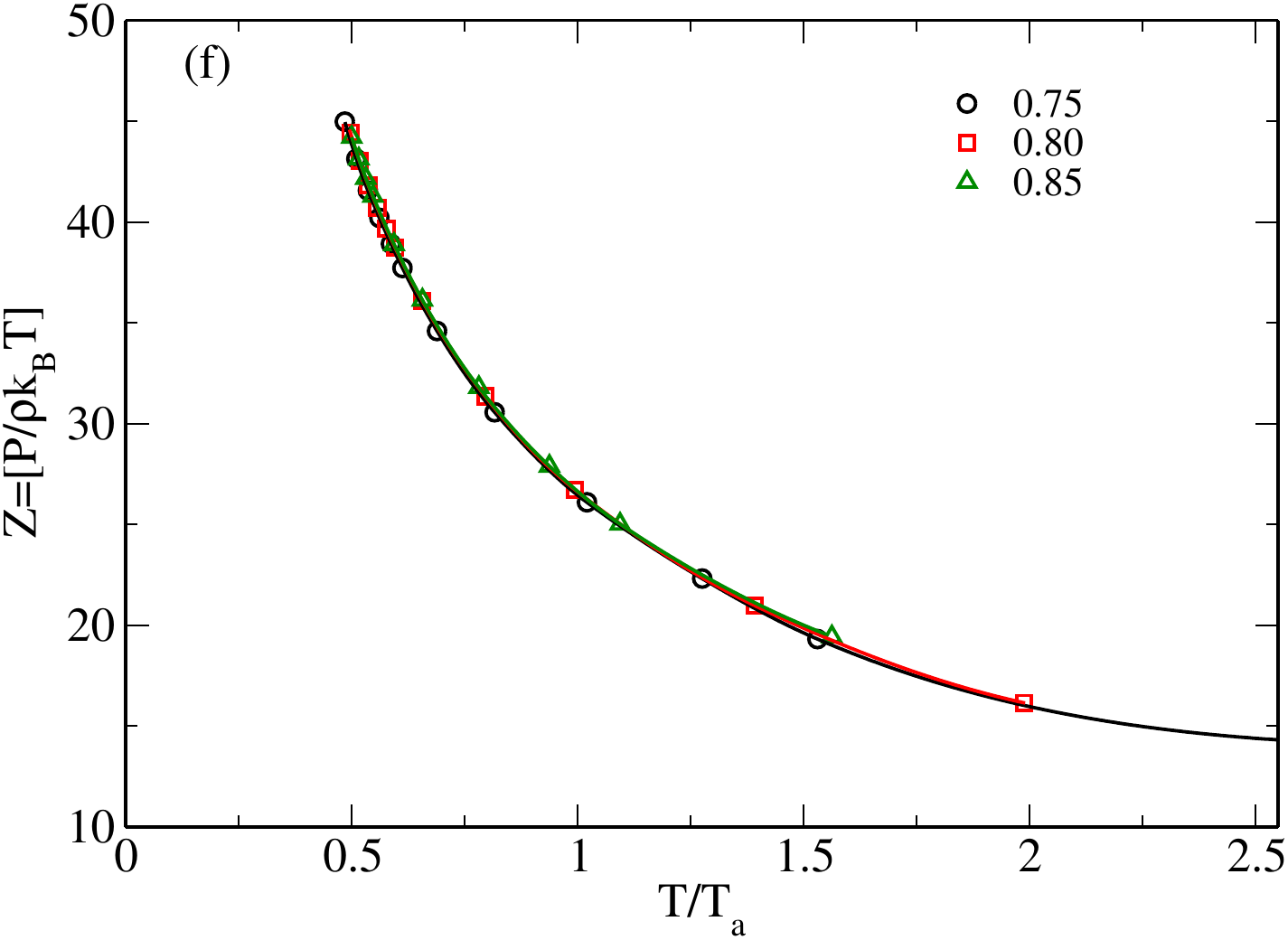}
    \end{minipage}

    \caption{Thermodynamic quantities for densities $\rho=0.75$, $0.80$, and $0.85$ collapse onto master curves when plotted as functions of $1/\Gamma$ (a–c) and $T/T_a$ (d–f). (a,d) scaled internal energy, $U/T_a$, from Fig.\ref{U_Cv_Z}(a); (b,e) specific heat, $C_V$, from Fig.\ref{U_Cv_Z}(b), fitted by $C_V=1.14,\Gamma^{1.7}+1.5$ in (b) and $C_V=1.73,(T/T_a)^{-0.42}+1.5$ in (e) (solid blue lines); and (c,f) reduced pressure, $Z=P/(\rho k_{\mathrm B}T)$, from Fig.~\ref{U_Cv_Z}(c).}
    \label{thermodynamic_Ta}
\end{figure*}

In Fig.~\ref{particle}, we plot number of bonded $b$-particles $[n_{1}^{(b)}(T)]$, $m$-particles $[n_{1}^{(m)}(T)]$ and $s$-particles $[n_{1}^{(s)}(T)]$ occupying the first shell as a function of inverse temperature $(1/T)$ at density $\rho=0.80$. The figure shows how temperature affects the number of particles of different kind in a shell. At high temperatures, the numbers of $b$-particles and $m$-particles are nearly the same, while $s$-particles constitute a smaller fraction than $m$-particles. Upon cooling, $n_1^{(m)}(T)$ remains nearly constant, whereas $n_1^{(s)}(T)$ increases slowly upto $T=T_a$. For $T<T_a$, $n_1^{(m)}(T)$ decreases, while $n_1^{(s)}(T)$ increases at an increasing rate at the cost of decreasing  both free and $m$-particles. This rate is expected to increase rapidly upon further lowering the temperature, resulting in a rapid increase in the number of $s$-particles.

In Fig.~\ref{energy}, we plot the dependence of $\beta E^{(s)}$ on inverse temperature $1/T$ for densities $\rho=0.75$, $0.80$ and $0.85$. It is evident that $\beta E^{(s)}$ increases rapidly on lowering the temperature for each density. This is due to the increase in the number of particles in the CRC. A relation $\beta E^{(s)}=b_{0} \left[1+n^{(s)}\right]^{\delta}$ with $b_{0}=0.18\pm0.02$ and $\delta=2.00\pm0.02$, reproduces the activation energy $\beta E^{(s)}$ very accurately. From the figure, it can be seen that the calculated numbers and the fitted expression for $\beta E^{(s)}$ are in very good agreement.
 
In Fig.~\ref{relaxation} we plot values of $\tau_{\alpha}$ as a function of $1/T$ for densities $\rho=0.75$, $0.80$, and $0.85$.  Here, the solid line is calculated from Eq.~\ref{tau_E} and the open circles represent simulation results obtained from the decay of the self-intermediate scattering function, as described in Sec.~\ref{Sec_relaxation}. The $\tau_\alpha$ values obtained from the theory show very good agreement with the simulation results for all three densities. As expected, at a given density, the relaxation time increases rapidly on cooling, indicating a dynamical slowdown, and at a given temperature, the relaxation time increases with the rise in density. 

\begin{figure}[b]
\includegraphics[width=0.4\textwidth,clip]{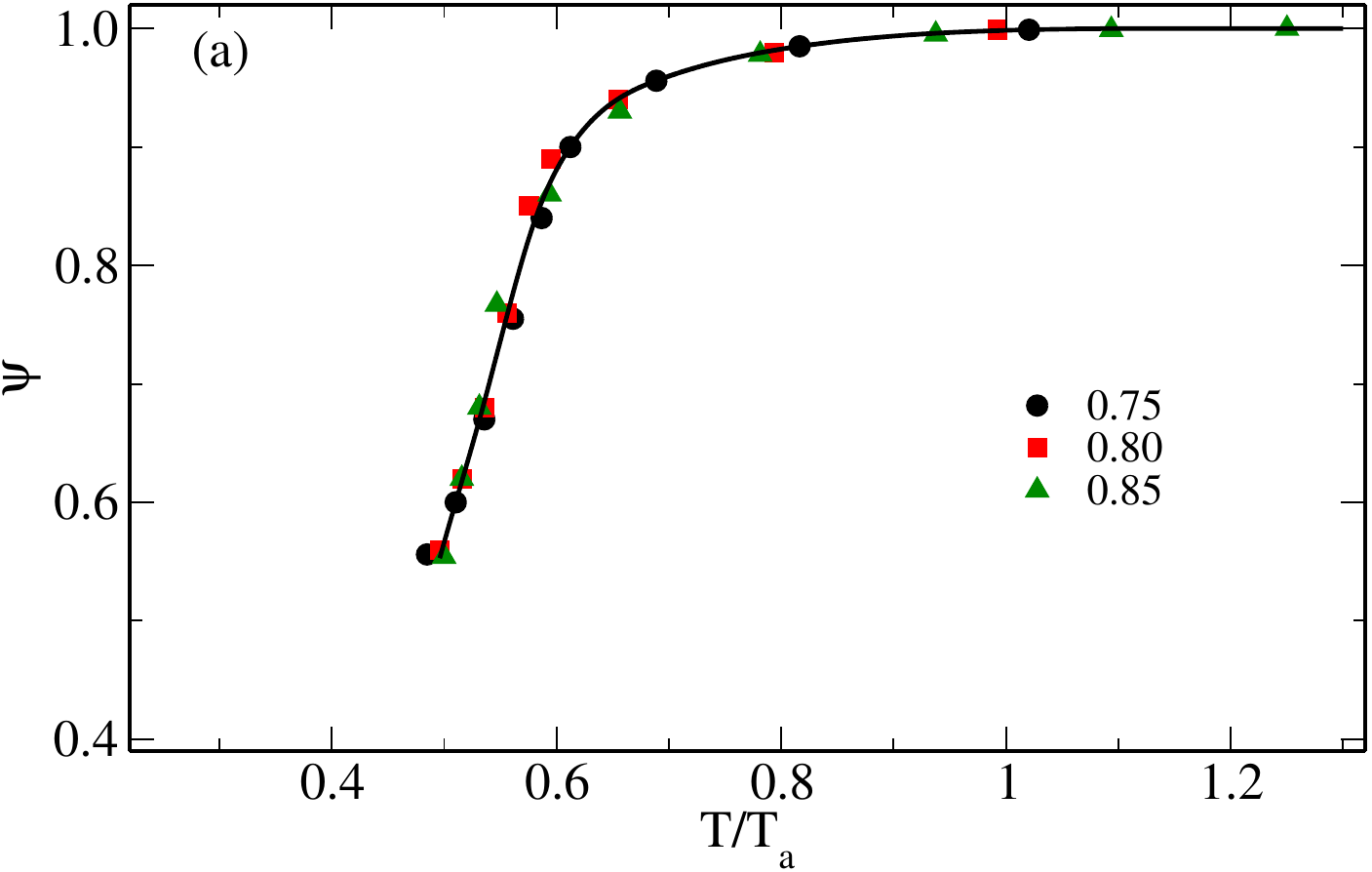}
\includegraphics[width=0.4\textwidth,clip]{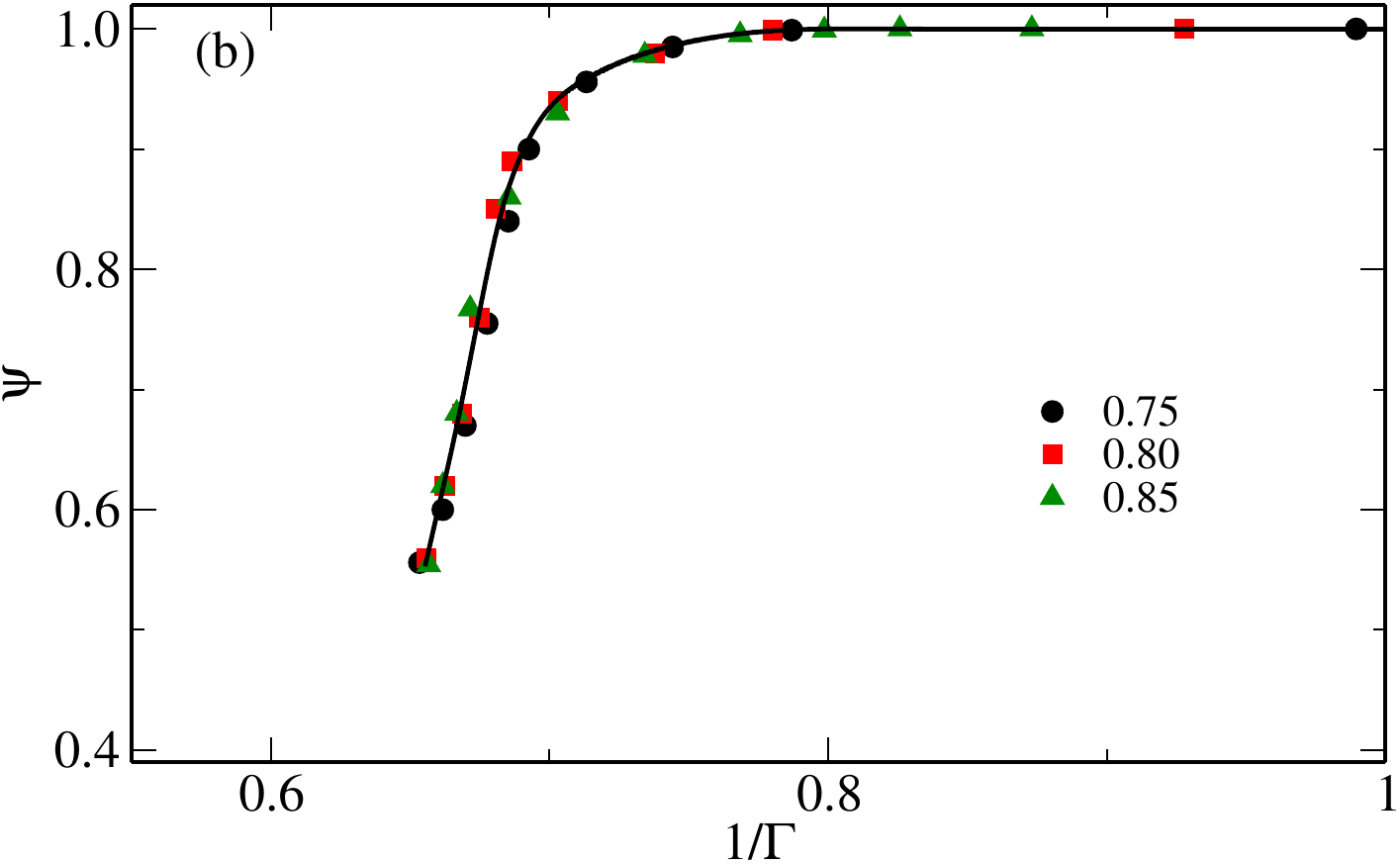}
\caption{Data collapse of $\psi$ at densities $\rho = 0.75$, $0.80$, and $0.85$. (a) $\psi$ plotted as a function of $T/T_a$. (b) $\psi$ plotted as a function of $1/\Gamma$. The line show Akima spline fits to the data.}
\label{psi_scaling}
\end{figure}

\section{Density-Temperature scaling \label{scaling}}

\begin{figure}[t]     
\includegraphics[width=0.45\textwidth,clip]{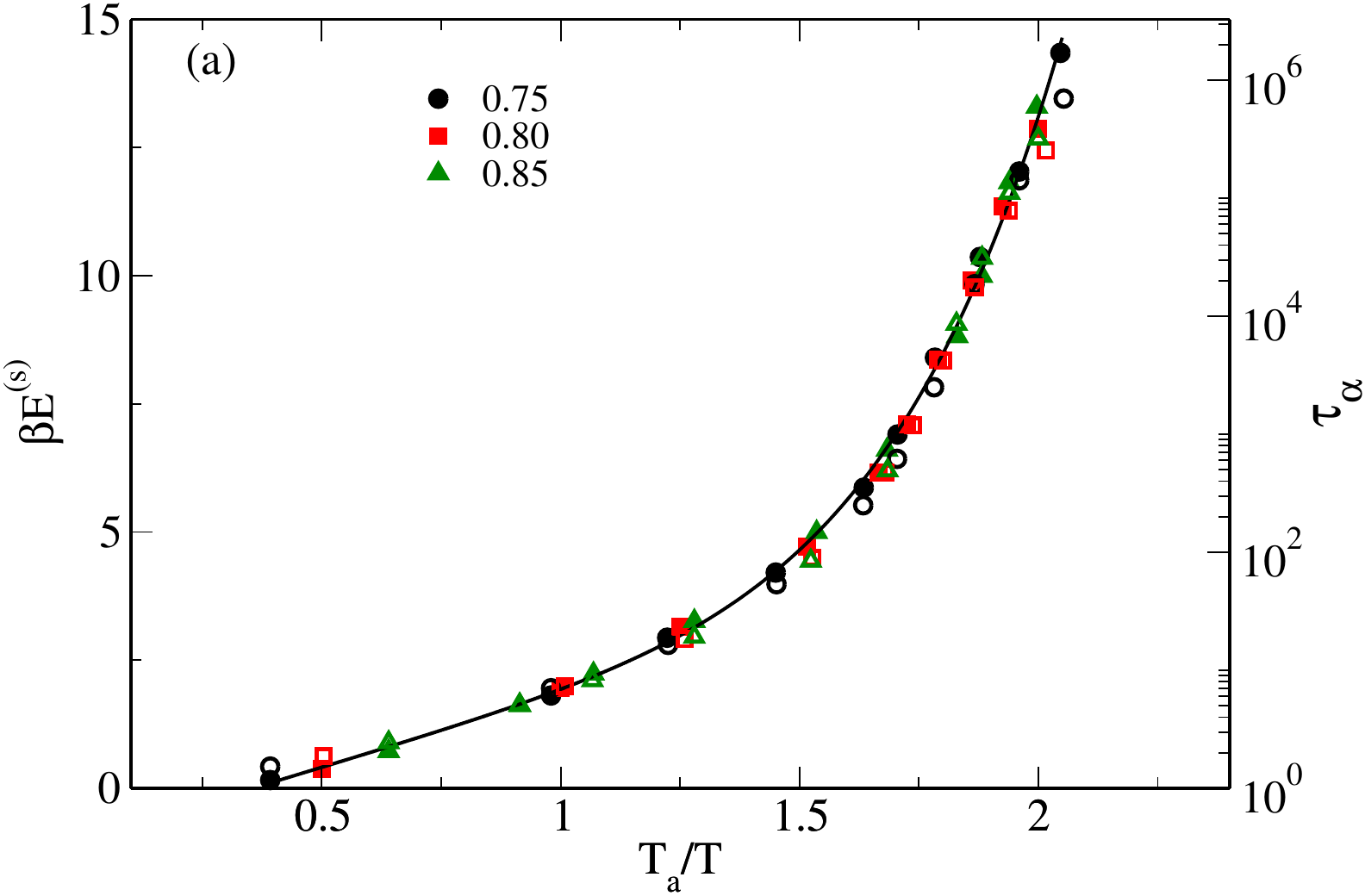}
\includegraphics[width=0.45\textwidth,clip]{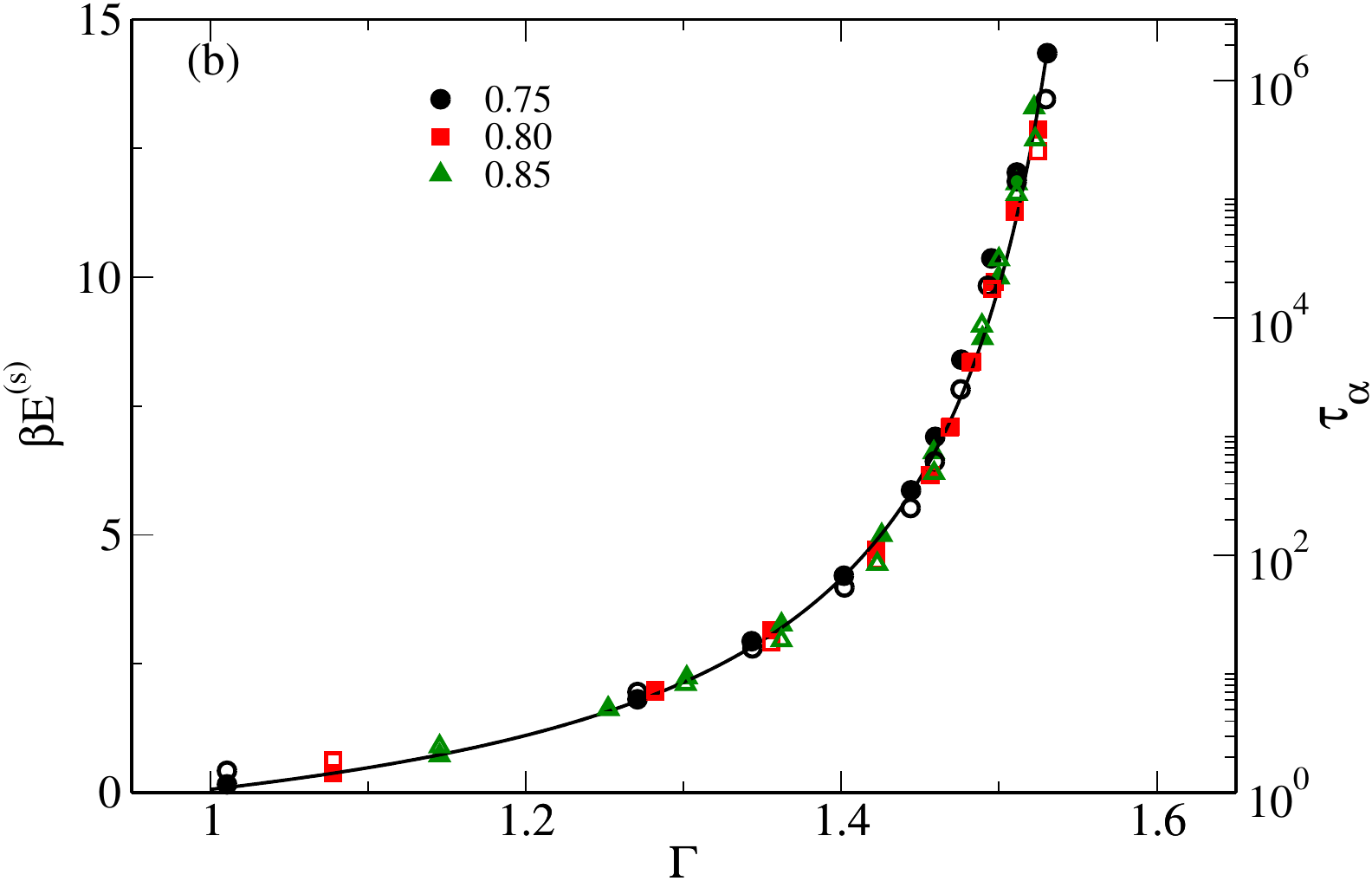}
    \caption{Collapse of $\beta E^{(s)}$ (LHS) and the structural relaxation time, $\tau_{\alpha}$ (RHS), for densities $\rho = 0.75$, $0.80$, and $0.85$. (a) $\beta E^{(s)}$ and $\tau_{\alpha}$ plotted as functions of $T_a/T$. (b) $\beta E^{(s)}$ and $\tau_{\alpha}$ plotted as functions of $\Gamma$. Solid symbols represent the calculated values of $\beta E^{(s)}$, while open symbols represent the $\tau_{\alpha}$ estimates obtained from MD simulations. The solid lines correspond to spline fits to the theoretical values.}
\label{relaxation_scaling}
\end{figure}

The excess thermodynamic quantities of IPL fluids are known to collapse on a (master) curve when plotted as functions of the parameter $\Gamma$ (see Eq.~\eqref{Gamma} for its definition) instead of the temperature $T$ \cite{berthier2011role,pedersen2010repulsive}. In Figs.~\ref{thermodynamic_Ta} (a-c), we show the master curves for the quantities $U$, $C_V$, and $Z$, respectively. These quantities also collapse onto master curves, as shown in Figs.~\ref{thermodynamic_Ta} (d-f), when $T$ is expressed in units of $T_a$. 

In Figs.~\ref{psi_scaling} (a) and (b), we plot $\psi(T)$ as a function of $T/T_a$ and $1/\Gamma$. In both cases, a very good collapse of the data at different densities is found. Similarly, when $\beta E^{(s)}(\rho,T)$, and $\tau_{\alpha}(\rho, T)$ are plotted as a function of $T_{a}/T$ and $\Gamma$, a very good collapse of data on master curves happens (see Figs.~\ref{relaxation_scaling} (a) and (b)); for the IPL fluids $T_{a}\propto\Gamma^{n/3}$. Here, solid symbols represent the calculated values of $\beta E^{(s)}$, while open symbols represent the $\tau_{\alpha}$ estimates obtained from MD simulations. The solid lines correspond to spline fits to the theoretical values.

Thermodynamic scaling through the variable $\rho^{\gamma}/T$ is found to be obeyed by a variety of liquids \cite{casalini2004thermodynamical, roland2005density, dyre2014hidden, veldhorst2014scaling}. A method which uses correlation between constant volume equilibrium fluctuations of potential energy $U_{p}(t)$ and virial $W=-\frac{1}{3}\sum_{i}\vec{r}_{i}\cdot\nabla_{\vec{r}_{i}} U_{p}(\vec{r}_{1},\vec{r}_{2},\ldots,\vec{r}_{N})$, where $U_{p}(t)$ is the total potential energy at time $t$ and $\vec{r}_{i}$ the position of particle $i$ at time $t$, was developed to determine $\gamma$ \cite{Pedersen_PhysRevLett_2008, bailey2008pressure, coslovich2009pressure}. In particular, $\gamma$ is determined from the slope of the correlation plot of time fluctuations of $U_{p}$ and $W$. Since the points representing the correlation in the plot are widely scattered, (see, e.g. Fig.~10 of Ref.~\cite{berthier2011role}) there is some arbitrariness in determining the slope. For example, for Kob–Andersen LJ (KALJ) liquid the reported value of $\gamma$ varies between 4.5 and 5.2 \cite{berthier2009nonperturbative, berthier2011role}. In contrast, the value of $\gamma$ for the fluid from the plot of  $T_{a}$ is found to be 4.757 \cite{PhysRevE.103.032611}. However, the analysis suggests a deep relationship between the $U_{p}$-$W$ correlations in time, fluctuations and the crossover temperature $T_{a}$. For the IPL fluids $\gamma=n/3$, where $n$ measures softness (see Eq.~\ref{eq:scaled-ipl-potential}) of the repulsion. This fact has been taken as an indication of predominance of the repulsive part of the pair interaction and led to search for an IPL potential which can reproduce the properties of the KALJ liquids \cite{pedersen2010repulsive}.

\begin{figure}[t]
\includegraphics[width=0.45\textwidth,clip]{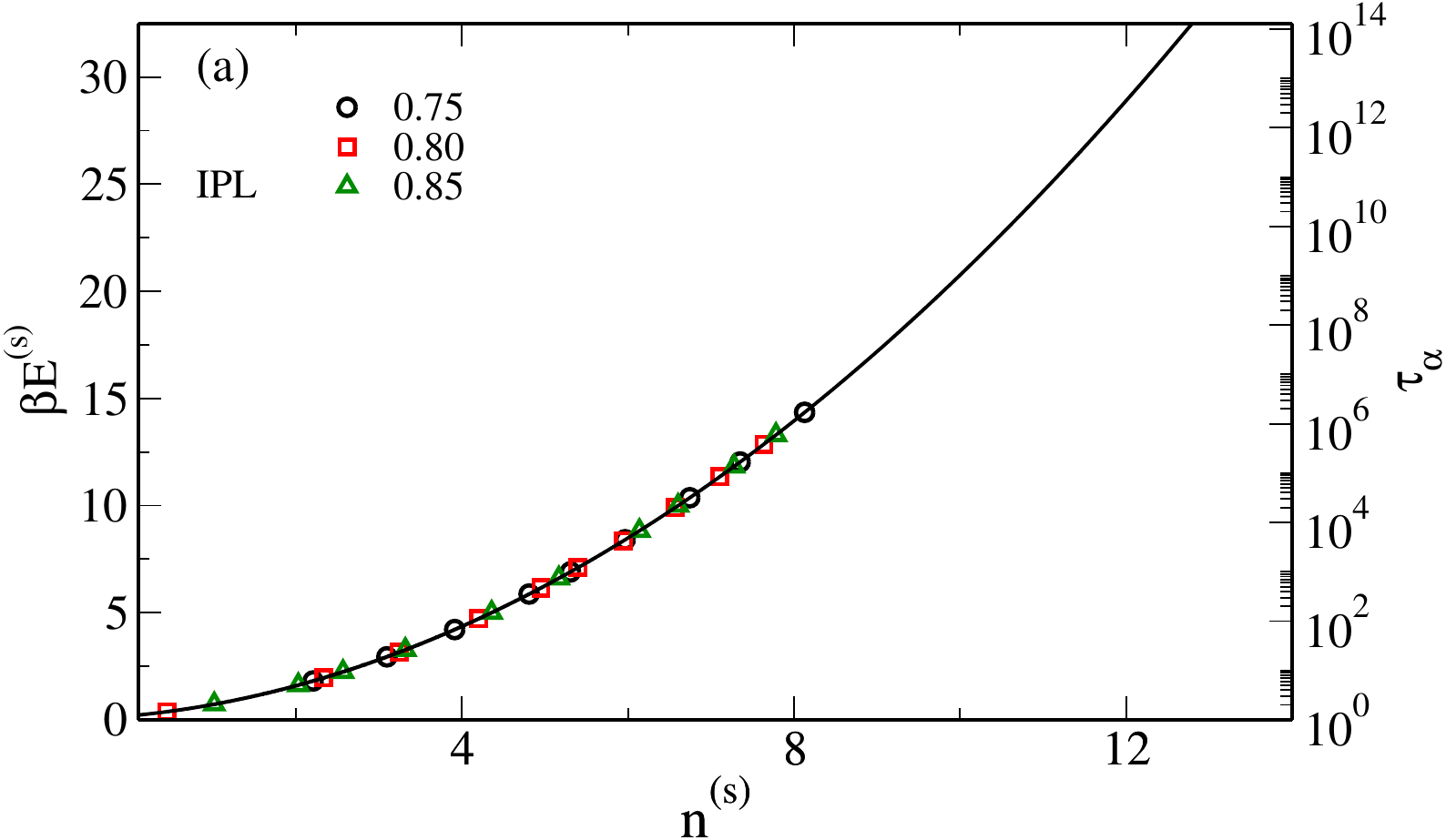}
\includegraphics[width=0.45\textwidth,clip]{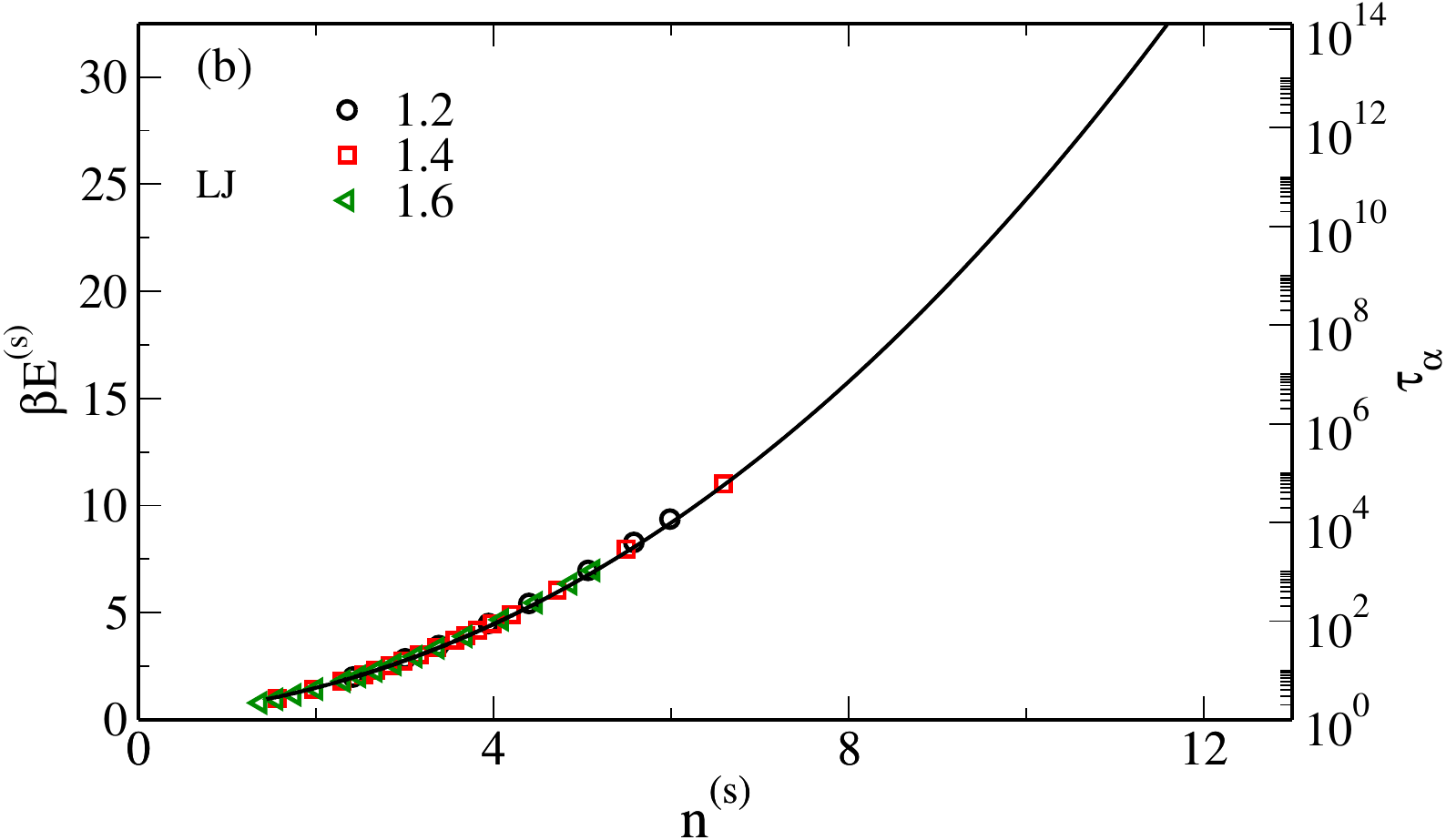}
\includegraphics[width=0.45\textwidth,clip]{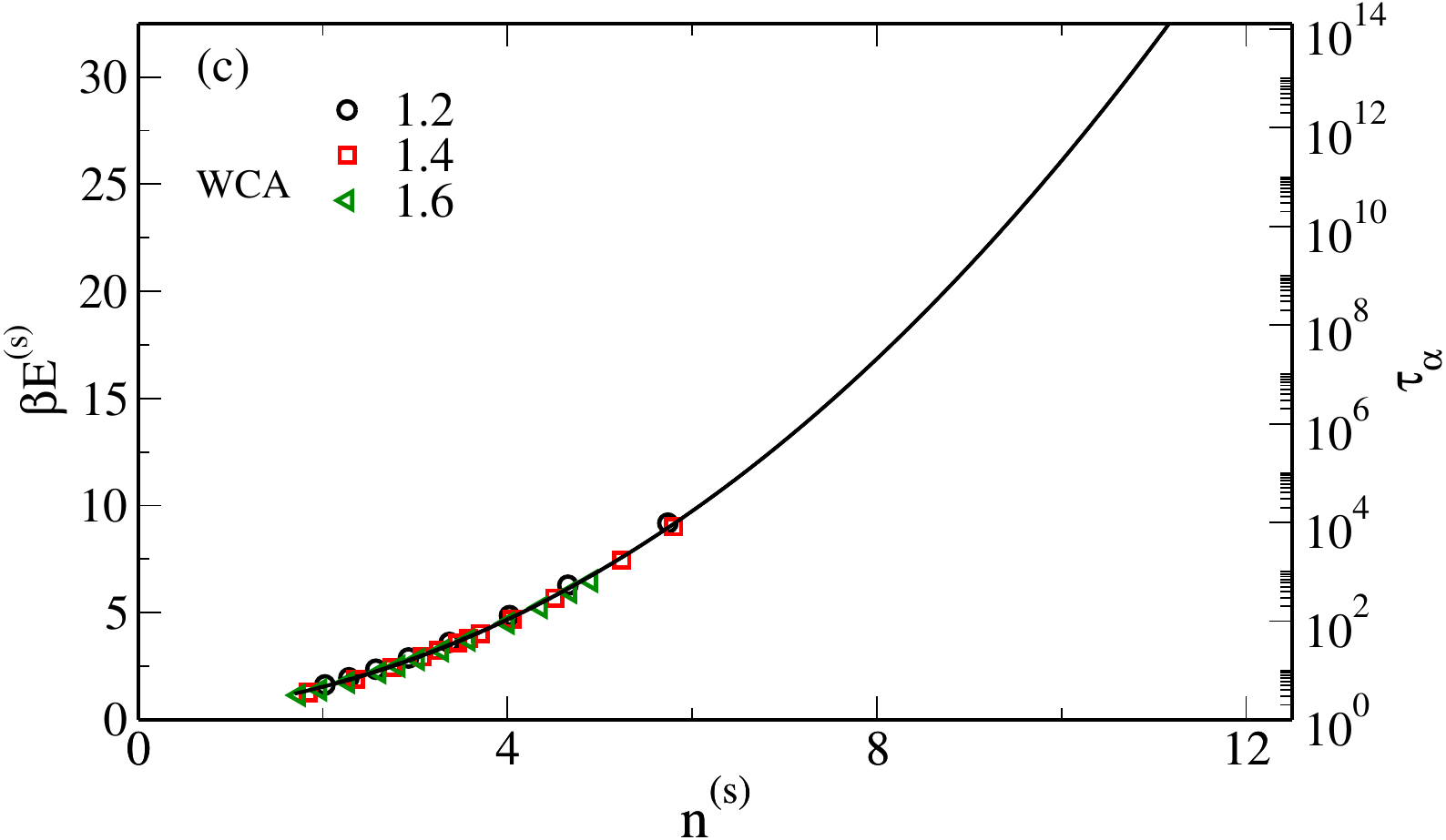}
\caption{Plots of $\beta E^{(s)}$ (LHS) and the structural relaxation time, $\tau_{\alpha}$ (RHS), as functions of $n^{(s)}$ for three densities (as indicated). Symbols represent the calculated values, while the solid lines correspond to equation fits upto $\tau_{\alpha}=10^{14}$ at the glass transition. (a) For IPL system, (b) For LJ system and (c) For WCA system.}
\label{LJ_WCA_ES_ns}
\end{figure}

\section{Summary and Conclusions}\label{conclusion}
The results presented above and in earlier papers \cite{PhysRevE.99.030101, PhysRevE.103.032611, PhysRevE.103.052105, PhysRevE.107.014119} for both the thermal and the athermal glass-formers show that the emergence and growth of CRC is primarily cause of dramatic slowing down of dynamics. The particles in a CRC form (nonchemical) ``stable bonds'' with the central particle and are distributed in coordination shells (cages) surrounding the central particle (Fig.~\ref{fig-1}). For an event of structural relaxation to take place, the CRC has to reorganize irreversibly; the energy involved in this rearrangement is the activation energy $\beta E^{(s)}$ of the relaxation. As the system is cooled, both the number of particles, $n^{(s)}$ and the energy of each bond in the CRC increase; the combined effect makes $\beta E^{(s)}$ to increase rapidly resulting into a super Arrhenius behavior.

The CRC is embedded at the center of a much larger cluster of $m$-particles which are loosely bonded with the central particle. The $m$-particles move individually or in group of a few particles on a timescale much smaller than $\tau_{\alpha}$ without affecting the structure of CRC. But when the CRC reorganizes at timescales commensurate with $\tau_{\alpha}$ by moving its particles, it triggers the reorganization of all $s$- and $m$-particles into a cluster of moveable particles \cite{PhysRevLett.96.057801}. The cluster that appears in simulations \cite{PhysRevLett.96.057801} and in experiments \cite{PhysRevLett.107.065704} that take part in relaxation is, in fact, the cluster of all bonded particles which is much larger than CRC and is relatively compact. We have, therefore, two ``length'' scales of relaxation; one defined by CRC and the other by the larger cluster of all bonded particles.

To estimate the number of $n^{(s)}$ particles in a CRC at the glass-transition temperature $T_g$, where $\tau_{\alpha}\sim10^{14}$, we use the relation $\beta E^{(s)}=b_0(1+n^{(s)})^{\delta}$ (see Fig.~\ref{energy}) and Eq.~\eqref{tau_E} for $\tau_{\alpha}$ to plot Fig.~\ref{relaxation}. The symbols in the figure represent the calculated values shown in Fig.~\ref{LJ_WCA_ES_ns} (a). From the figure, we note that the number of particles $(1+n^{(s)})$ in a CRC at $T_g$ is $\sim14$, which is a small number. To confirm this, we perform similar calculations for the KALJ and KAWCA systems using results given in Refs.~\cite{PhysRevE.103.032611, PhysRevE.103.052105}. Here, the $\beta E^{(s)}$ fitted parameters are $b_0=0.13$ and $\delta=2.15$ for the LJ system, and $b_0=0.13$ and $\delta=2.18$ for the WCA system. Results of these calculations are plotted in Figs.~\ref{LJ_WCA_ES_ns}~(b-c), which also show that the number of particles in a CRC is of the order of 14. These particles are confined at most within the first two shells of the effective potential (see Fig.~\ref{fig-1}). However, the number of $m$ particles in the cluster at which centre the CRC is embedded is in hundreds.

The crossover temperature, which separates the low-temperature dynamics from that of high temperature and is regarded as a characteristic feature of thermal glass formers, manifests itself in different properties of the system. For example, as shown in Sec.~\ref{sec:model_method}, it appears in the time dependence of the quantity $F_s(q,t)$ and in the inherent energy. The most noticeable separation is seen in $\psi(T)$, where, for $T<T_a$, it starts decreasing sharply from its high-temperature value of 1. The crossover of $\psi(T)$ from high temperature to low temperature is in one-to-one correspondence with $\tau_{\alpha}$. The quantity $\psi(T)$ measures the effect of fluctuations embedded in the bath on the formation of a CRC. We conclude with a remark that the CRC which is determined from the static pair correlation function and the fluctuations embedded in the system is probably the sole player in dramatic slowing down of dynamics in the fragile glass formers.\\

\section*{Acknowledgment}
AS and SLS acknowledge the generous financial support received from BHU under the IoE scheme. AS is also grateful for funding from the European Union through the Horizon Europe ERC Grant No.~101043968, ``Multimech’’. SLS also acknowledges financial support from CSIR, New Delhi. VV gratefully acknowledges the high-performance computing facilities at the Institute of Mathematical Sciences Chennai for providing computing time.

\subsection*{Author Contributions}
A.S. and V.V. contributed equally to this work.

\bibliography{article}

@article{berthier2011theoretical,
  title = {Theoretical perspective on the glass transition and amorphous materials},
  author = {Berthier, Ludovic and Biroli, Giulio},
  journal = {Rev. Mod. Phys.},
  volume = {83},
  issue = {2},
  pages = {587--645},
  numpages = {0},
  year = {2011},
  month = {Jun},
  publisher = {American Physical Society},
  doi = {10.1103/RevModPhys.83.587}
}

@Article{Schoenholz_2016,
author={Schoenholz, S. S.
and Cubuk, E. D.
and Sussman, D. M.
and Kaxiras, E.
and Liu, A. J.},
title={A structural approach to relaxation in glassy liquids},
journal={Nature Physics},
year={2016},
month={May},
day={01},
volume={12},
number={5},
pages={469-471},
issn={1745-2481},
doi={10.1038/nphys3644}
}

@article{tarjus2011overview,
  title={An overview of the theories of the glass transition},
  author={Tarjus, Gilles},
  journal={Dynamical Heterogeneities in Glasses, Colloids, and Granular Media},
  volume={150},
  pages={39},
  year={2011},
  publisher={Oxford University Press}
}

@article{biroli2013perspective,
  title={Perspective: The glass transition},
  author={Biroli, Giulio and Garrahan, Juan P},
  journal={The Journal of chemical physics},
  volume={138},
  number={12},
  year={2013},
  publisher={AIP Publishing}
}

@article{Mallamace_2010,
author = {Francesco Mallamace  and Caterina Branca  and Carmelo Corsaro  and Nancy Leone  and Jeroen Spooren  and Sow-Hsin Chen  and H. Eugene Stanley },
title = {Transport properties of glass-forming liquids suggest that dynamic crossover temperature is as important as the glass transition temperature},
journal = {Proceedings of the National Academy of Sciences},
volume = {107},
number = {52},
pages = {22457-22462},
year = {2010},
doi = {10.1073/pnas.1015340107}}

@article{EdigerHarrowell_2012,
  author  = {Ediger, Mark D. and Harrowell, Peter},
  title   = {Perspective: Supercooled Liquids and Glasses},
  journal = {The Journal of Chemical Physics},
  volume  = {137},
  pages   = {080901},
  year    = {2012},
  doi     = {10.1063/1.4747326}
}

@article{BouchaudBiroli2004,
  author  = {Bouchaud, Jean-Philippe and Biroli, Giulio},
  title   = {On the {Adam--Gibbs--Kirkpatrick--Thirumalai--Wolynes} Scenario for the Viscosity Increase in Glasses},
  journal = {The Journal of Chemical Physics},
  volume  = {121},
  pages   = {7347--7354},
  year    = {2004},
  doi     = {10.1063/1.1796231}
}

@article{MontanariSemerjian2006,
  author  = {Montanari, Andrea and Semerjian, Guilhem},
  title   = {Rigorous Inequalities between Length and Time Scales in Glassy Systems},
  journal = {Journal of Statistical Physics},
  volume  = {125},
  number  = {1},
  pages   = {23--54},
  year    = {2006},
  doi     = {10.1007/s10955-006-9175-y}
}

@article{Kawasaki2007,
  author  = {Kawasaki, Takeshi and Araki, Takeaki and Tanaka, Hajime},
  title   = {Correlation between Dynamic Heterogeneity and Medium-Range Order in Two-Dimensional Glass-Forming Liquids},
  journal = {Physical Review Letters},
  volume  = {99},
  pages   = {215701},
  year    = {2007},
  doi     = {10.1103/PhysRevLett.99.215701}
}

@article{KivelsonTarjus2008,
  author  = {Kivelson, Steven A. and Tarjus, Gilles},
  title   = {In Search of a Theory of Supercooled Liquids},
  journal = {Nature Materials},
  volume  = {7},
  pages   = {831--833},
  year    = {2008},
  doi     = {10.1038/nmat2304}
}

@article{Biroli2008,
  author  = {Biroli, Giulio and Bouchaud, Jean-Philippe and Cavagna, Andrea and Grigera, Tom{\'a}s S. and Verrocchio, Paolo},
  title   = {Thermodynamic Signature of Growing Amorphous Order in Glass-Forming Liquids},
  journal = {Nature Physics},
  volume  = {4},
  pages   = {771--775},
  year    = {2008},
  doi     = {10.1038/nphys1050}
}

@article{Mosayebi2010,
  author  = {Mosayebi, Majid and Del Gado, Emanuela and Ilg, Patrick and {\"O}ttinger, Hans Christian},
  title   = {Probing a Critical Length Scale at the Glass Transition},
  journal = {Physical Review Letters},
  volume  = {104},
  pages   = {205704},
  year    = {2010},
  doi     = {10.1103/PhysRevLett.104.205704}
}

@article{KurchanLevine2011,
  author  = {Kurchan, Jorge and Levine, Dov},
  title   = {Order in Glassy Systems},
  journal = {Journal of Physics A: Mathematical and Theoretical},
  volume  = {44},
  number  = {3},
  pages   = {035001},
  year    = {2011},
  doi     = {10.1088/17518113/44/3/035001}
}

@article{SaussetLevine2011,
  author  = {Sausset, Fran{\c{c}}ois and Levine, Dov},
  title   = {Characterizing Order in Amorphous Systems},
  journal = {Physical Review Letters},
  volume  = {107},
  pages   = {045501},
  year    = {2011},
  doi     = {10.1103/PhysRevLett.107.045501}
}

@article{Hocky2012,
  author  = {Hocky, Glen M. and Markland, Thomas E. and Reichman, David R.},
  title   = {Growing Point-to-Set Length Scale Correlates with Growing Relaxation Times in Model Supercooled Liquids},
  journal = {Physical Review Letters},
  volume  = {108},
  pages   = {225506},
  year    = {2012},
  doi     = {10.1103/PhysRevLett.108.225506}
}

@article{CammarotaBiroli2012,
  author  = {Cammarota, Chiara and Biroli, Giulio},
  title   = {Patch-Repetition Correlation Length in Glassy Systems},
  journal = {Europhysics Letters},
  volume  = {98},
  number  = {3},
  pages   = {36005},
  year    = {2012},
  doi     = {10.1209/0295-5075/98/36005}
}

@article{Berthier2017,
  author  = {Berthier, Ludovic and Charbonneau, Patrick and Coslovich, Daniele and Ninarello, Andrea and Ozawa, Misaki and Yaida, Sho},
  title   = {Configurational Entropy Measurements in Extremely Supercooled Liquids that Break the Glass Ceiling},
  journal = {Proceedings of the National Academy of Sciences of the United States of America},
  volume  = {114},
  number  = {43},
  pages   = {11356--11361},
  year    = {2017},
  doi     = {10.1073/pnas.1706860114}
}

@article{NissHecksher2018,
  author  = {Niss, Kristine and Hecksher, Tina},
  title   = {Perspective: Searching for Simplicity Rather than
             Universality in Glass-Forming Liquids},
  journal = {The Journal of Chemical Physics},
  volume  = {149},
  number  = {23},
  pages   = {230901},
  year    = {2018},
  doi     = {10.1063/1.5048093}
}

@article{BerthierReichman2023,
  author  = {Berthier, Ludovic and Reichman, David R.},
  title   = {Modern Computational Studies of the Glass Transition},
  journal = {Nature Reviews Physics},
  volume  = {5},
  pages   = {102--116},
  year    = {2023},
  doi     = {10.1038/s42254-022-00548-x}
}

@article{DyreEdiger2026,
  author  = {Dyre, Jeppe C. and Ediger, Mark D.},
  title   = {Physics and Chemistry Perspectives on Three Unsolved
             Problems in Glass Science},
  journal = {Nature Reviews Physics},
  volume  = {8},
  pages   = {383--396},
  year    = {2026},
  doi     = {10.1038/s42254-026-00940-x}
}

@article{Chattoraj_2020,
  title = {Role of Attractive Forces in the Relaxation Dynamics of Supercooled Liquids},
  author = {Chattoraj, Joyjit and Ciamarra, Massimo Pica},
  journal = {Phys. Rev. Lett.},
  volume = {124},
  issue = {2},
  pages = {028001},
  numpages = {5},
  year = {2020},
  month = {Jan},
  publisher = {American Physical Society},
  doi = {10.1103/PhysRevLett.124.028001}
}

@article{Schweizer_2015,
  title = {Microscopic Theory for the Role of Attractive Forces in the Dynamics of Supercooled Liquids},
  author = {Dell, Zachary E. and Schweizer, Kenneth S.},
  journal = {Phys. Rev. Lett.},
  volume = {115},
  issue = {20},
  pages = {205702},
  numpages = {5},
  year = {2015},
  month = {Nov},
  publisher = {American Physical Society},
  doi = {10.1103/PhysRevLett.115.205702}}

@article{karmakar2009growing,
  title={Growing length and time scales in glass-forming liquids},
  author={Karmakar, Smarajit and Dasgupta, Chandan and Sastry, Srikanth},
  journal={Proceedings of the National Academy of Sciences},
  volume={106},
  number={10},
  pages={3675--3679},
  year={2009},
  publisher={National Academy of Sciences}
}

@article{PhysRevE.99.030101,
  title = {Super-Arrhenius behavior of molecular glass formers},
  author = {Singh, Ankit and Singh, Yashwant},
  journal = {Phys. Rev. E},
  volume = {99},
  issue = {3},
  pages = {030101},
  numpages = {4},
  year = {2019},
  month = {Mar},
  publisher = {American Physical Society},
  doi = {10.1103/PhysRevE.99.030101},
  url = {https://link.aps.org/doi/10.1103/PhysRevE.99.030101}
}

@article{PhysRevE.103.032611,
  title = {Emergence of cooperatively reorganizing cluster and super-Arrhenius dynamics of fragile supercooled liquids},
  author = {Singh, Ankit and Bhattacharyya, Sarika Maitra and Singh, Yashwant},
  journal = {Phys. Rev. E},
  volume = {103},
  issue = {3},
  pages = {032611},
  numpages = {9},
  year = {2021},
  month = {Mar},
  publisher = {American Physical Society},
  doi = {10.1103/PhysRevE.103.032611},
  url = {https://link.aps.org/doi/10.1103/PhysRevE.103.032611}
}

@article{PhysRevE.103.052105,
  title = {How attractive and repulsive interactions affect structure ordering and dynamics of glass-forming liquids},
  author = {Singh, Ankit and Singh, Yashwant},
  journal = {Phys. Rev. E},
  volume = {103},
  issue = {5},
  pages = {052105},
  numpages = {8},
  year = {2021},
  month = {May},
  publisher = {American Physical Society},
  doi = {10.1103/PhysRevE.103.052105},
  url = {https://link.aps.org/doi/10.1103/PhysRevE.103.052105}
}

@article{PhysRevE.107.014119,
  title = {Structure ordering and glass transition in size-asymmetric ternary mixtures of hard spheres: Variation from fragile to strong glasses},
  author = {Singh, Ankit and Singh, Yashwant},
  journal = {Phys. Rev. E},
  volume = {107},
  issue = {1},
  pages = {014119},
  numpages = {14},
  year = {2023},
  month = {Jan},
  publisher = {American Physical Society},
  doi = {10.1103/PhysRevE.107.014119},
  url = {https://link.aps.org/doi/10.1103/PhysRevE.107.014119}
}

@article{Hoover,
    author = {Hoover, William G. and Ross, Marvin and Johnson, Keith W. and Henderson, Douglas and Barker, John A. and Brown, Bryan C.},
    title = "{Soft‐Sphere Equation of State}",
    journal = {The Journal of Chemical Physics},
    volume = {52},
    number = {10},
    pages = {4931-4941},
    year = {1970},
    month = {05},
    issn = {0021-9606},
    doi = {10.1063/1.1672728},
    url = {https://doi.org/10.1063/1.1672728},
}

@article{WHoover,
    author = {Hoover, William G. and Gray, Steven G. and Johnson, Keith W.},
    title = "{Thermodynamic Properties of the Fluid and Solid Phases for Inverse Power Potentials}",
    journal = {The Journal of Chemical Physics},
    volume = {55},
    number = {3},
    pages = {1128-1136},
    year = {1971},
    month = {08},
    issn = {0021-9606},
    doi = {10.1063/1.1676196},
    url = {https://doi.org/10.1063/1.1676196},
    
}

@article{SINGH1991351,
title = {Density-functional theory of freezing and properties of the ordered phase},
journal = {Physics Reports},
volume = {207},
number = {6},
pages = {351-444},
year = {1991},
issn = {0370-1573},
doi = {https://doi.org/10.1016/0370-1573(91)90097-6},
url = {https://www.sciencedirect.com/science/article/pii/0370157391900976},
author = {Yashwant Singh}
}

@book{Hansen,
title = {},
booktitle = {Theory of Simple Liquids},
publisher = {Academic Press},
edition = {Third Edition},
address = {Burlington, VT},
pages = {},
year = {2006},
isbn = {978-0-12-370535-8},
doi = {https://doi.org/10.1016/B978-012370535-8/50000-8},
url = {https://www.sciencedirect.com/science/article/pii/B9780123705358500008},
author = {Jean-Pierre Hansen and Ian R. McDonald}
}

@article{Adam,
    author = {Adam, Gerold and Gibbs, Julian H.},
    title = "{On the Temperature Dependence of Cooperative Relaxation Properties in Glass‐Forming Liquids}",
    journal = {The Journal of Chemical Physics},
    volume = {43},
    number = {1},
    pages = {139-146},
    year = {1965},
    month = {07},
    issn = {0021-9606},
    doi = {10.1063/1.1696442},
    url = {https://doi.org/10.1063/1.1696442},
}

@article{Bouchaud,
    author = {Bouchaud, Jean-Philippe and Biroli, Giulio},
    title = "{On the Adam-Gibbs-Kirkpatrick-Thirumalai-Wolynes scenario for the viscosity increase in glasses}",
    journal = {The Journal of Chemical Physics},
    volume = {121},
    number = {15},
    pages = {7347-7354},
    year = {2004},
    month = {10},
    issn = {0021-9606},
    doi = {10.1063/1.1796231},
    url = {https://doi.org/10.1063/1.1796231},
}

@article{PhysRevLett.96.057801,
  title = {Democratic Particle Motion for Metabasin Transitions in Simple Glass Formers},
  author = {Appignanesi, G. A. and Rodr\'{\i}guez Fris, J. A. and Montani, R. A. and Kob, W.},
  journal = {Phys. Rev. Lett.},
  volume = {96},
  issue = {5},
  pages = {057801},
  numpages = {4},
  year = {2006},
  month = {Feb},
  publisher = {American Physical Society},
  doi = {10.1103/PhysRevLett.96.057801},
  url = {https://link.aps.org/doi/10.1103/PhysRevLett.96.057801}
}

@article{PhysRevLett.107.065704,
  title = {Experimental Verification of Rapid, Sporadic Particle Motions by Direct Imaging of Glassy Colloidal Systems},
  author = {Fris, J. Ariel Rodriguez and Appignanesi, Gustavo A. and Weeks, Eric R.},
  journal = {Phys. Rev. Lett.},
  volume = {107},
  issue = {6},
  pages = {065704},
  numpages = {5},
  year = {2011},
  month = {Aug},
  publisher = {American Physical Society},
  doi = {10.1103/PhysRevLett.107.065704},
  url = {https://link.aps.org/doi/10.1103/PhysRevLett.107.065704}
}

@article{pedersen2010repulsive,
  title = {Repulsive Reference Potential Reproducing the Dynamics of a Liquid with Attractions},
  author = {Pedersen, Ulf R. and Schr\o{}der, Thomas B. and Dyre, Jeppe C.},
  journal = {Phys. Rev. Lett.},
  volume = {105},
  issue = {15},
  pages = {157801},
  numpages = {4},
  year = {2010},
  month = {Oct},
  publisher = {American Physical Society},
  doi = {10.1103/PhysRevLett.105.157801}
}

@article{berthier2009nonperturbative,
  title = {Nonperturbative Effect of Attractive Forces in Viscous Liquids},
  author = {Berthier, Ludovic and Tarjus, Gilles},
  journal = {Phys. Rev. Lett.},
  volume = {103},
  issue = {17},
  pages = {170601},
  numpages = {4},
  year = {2009},
  month = {Oct},
  publisher = {American Physical Society},
  doi = {10.1103/PhysRevLett.103.170601}
}

@article{berthier2011role,
  title={The role of attractive forces in viscous liquids},
  author={Berthier, Ludovic and Tarjus, Gilles},
  journal={The Journal of chemical physics},
  volume={134},
  number={21},
  year={2011},
  publisher={AIP Publishing}
}

@article{bernu1985molecular,
  title={A molecular dynamics study of the glass transition in binary mixtures of soft spheres},
  author={Bernu, B and Hiwatari, Y and Hansen, JP},
  journal={Journal of Physics C: Solid State Physics},
  volume={18},
  number={14},
  pages={L371--L376},
  year={1985}
}

@article{bernu1987soft,
  title={Soft-sphere model for the glass transition in binary alloys: Pair structure and self-diffusion},
  author={Bernu, B and Hansen, JP and Hiwatari, Y and Pastore, GIORGIO},
  journal={Physical Review A},
  volume={36},
  number={10},
  pages={4891},
  year={1987},
  publisher={APS}
}

@article{thompson2022lammps,
  title={LAMMPS-a flexible simulation tool for particle-based materials modeling at the atomic, meso, and continuum scales},
  author={Thompson, Aidan P and Aktulga, H Metin and Berger, Richard and Bolintineanu, Dan S and Brown, W Michael and Crozier, Paul S and In't Veld, Pieter J and Kohlmeyer, Axel and Moore, Stan G and Nguyen, Trung Dac and others},
  journal={Computer physics communications},
  volume={271},
  pages={108171},
  year={2022},
  publisher={Elsevier}
}

@article{vaibhav2022finite,
  title={Finite-size effects in the diffusion dynamics of a glass-forming binary mixture with large size ratio},
  author={Vaibhav, Vinay and Horbach, J{\"u}rgen and Chaudhuri, Pinaki},
  journal={The Journal of Chemical Physics},
  volume={156},
  number={24},
  year={2022},
  publisher={AIP Publishing}
}

@article{sastry2000evaluation,
  title={Evaluation of the configurational entropy of a model liquid from computer simulations},
  author={Sastry, Srikanth},
  journal={Journal of Physics: Condensed Matter},
  volume={12},
  number={29},
  pages={6515--6523},
  year={2000}
}

@article{das2022crossover,
  title={Crossover in dynamics in the Kob-Andersen binary mixture glass-forming liquid},
  author={Das, Pallabi and Sastry, Srikanth},
  journal={Journal of Non-Crystalline Solids: X},
  volume={14},
  pages={100098},
  year={2022},
  publisher={Elsevier}
}

@article{berthier2019configurational,
  title={Configurational entropy of glass-forming liquids},
  author={Berthier, Ludovic and Ozawa, Misaki and Scalliet, Camille},
  journal={The Journal of chemical physics},
  volume={150},
  number={16},
  year={2019},
  publisher={AIP Publishing}
}

@book{binder2011glassy,
  title={Glassy materials and disordered solids: An introduction to their statistical mechanics},
  author={Binder, Kurt and Kob, Walter},
  year={2011},
  publisher={World scientific}
}

@article{coslovich2025freezing,
  title={Freezing, melting, and the onset of glassiness in binary mixtures},
  author={Coslovich, Daniele and Galliano, Leonardo and Costigliola, Lorenzo},
  journal={The Journal of Chemical Physics},
  volume={162},
  number={6},
  year={2025},
  publisher={AIP Publishing}
}

@article{casalini2004thermodynamical,
  title={Thermodynamical scaling of the glass transition dynamics},
  author={Casalini, R and Roland, CM},
  journal={Physical Review E—Statistical, Nonlinear, and Soft Matter Physics},
  volume={69},
  number={6},
  pages={062501},
  year={2004},
  publisher={APS}
}

@article{banerjee2017determination,
  title={Determination of onset temperature from the entropy for fragile to strong liquids},
  author={Banerjee, Atreyee and Nandi, Manoj Kumar and Sastry, Srikanth and Maitra Bhattacharyya, Sarika},
  journal={The Journal of chemical physics},
  volume={147},
  number={2},
  year={2017},
  publisher={AIP Publishing}
}

@article{bailey2008pressure,
  title={Pressure-energy correlations in liquids. I. Results from computer simulations},
  author={Bailey, Nicholas P and Pedersen, Ulf R and Gnan, Nicoletta and Schr{\o}der, Thomas B and Dyre, Jeppe C},
  journal={The Journal of chemical physics},
  volume={129},
  number={18},
  year={2008},
  publisher={AIP Publishing}
}

@article{tong2020role,
  title={Role of attractive interactions in structure ordering and dynamics of glass-forming liquids},
  author={Tong, Hua and Tanaka, Hajime},
  journal={Physical review letters},
  volume={124},
  number={22},
  pages={225501},
  year={2020},
  publisher={APS}
}

@article{landes2020attractive,
  title={Attractive versus truncated repulsive supercooled liquids: The dynamics is encoded in the pair correlation function},
  author={Landes, Fran{\c{c}}ois P and Biroli, Giulio and Dauchot, Olivier and Liu, Andrea J and Reichman, David R},
  journal={Physical Review E},
  volume={101},
  number={1},
  pages={010602},
  year={2020},
  publisher={APS}
}

@article{debenedetti2001supercooled,
  title={Supercooled liquids and the glass transition},
  author={Debenedetti, Pablo G and Stillinger, Frank H},
  journal={Nature},
  volume={410},
  number={6825},
  pages={259--267},
  year={2001},
  publisher={Nature Publishing Group UK London}
}

@book{gotze2009complex,
  title={Complex dynamics of glass-forming liquids: A mode-coupling theory},
  author={G{\"o}tze, Wolfgang},
  volume={143},
  year={2009},
  publisher={Oxford University Press}
}

@article{angell1995formation,
  title={Formation of glasses from liquids and biopolymers},
  author={Angell, C Austen},
  journal={Science},
  volume={267},
  number={5206},
  pages={1924--1935},
  year={1995},
  publisher={American Association for the Advancement of Science}
}

@article{donati1998stringlike,
  title={Stringlike cooperative motion in a supercooled liquid},
  author={Donati, Claudio and Douglas, Jack F and Kob, Walter and Plimpton, Steven J and Poole, Peter H and Glotzer, Sharon C},
  journal={Physical review letters},
  volume={80},
  number={11},
  pages={2338},
  year={1998},
  publisher={APS}
}

@article{banerjee2014role,
  title={Role of structure and entropy in determining differences in dynamics for glass formers with different interaction potentials},
  author={Banerjee, Atreyee and Sengupta, Shiladitya and Sastry, Srikanth and Bhattacharyya, Sarika Maitra},
  journal={Physical review letters},
  volume={113},
  number={22},
  pages={225701},
  year={2014},
  publisher={APS}
}

@article{rosenfeld1998density,
  title={Density functional theory and the asymptotic high density expansion of the free energy of classical solids and fluids},
  author={Rosenfeld, Yaakov and Tarazona, Pedro},
  journal={Molecular Physics},
  volume={95},
  number={2},
  pages={141--150},
  year={1998},
  publisher={Taylor \& Francis}
}

@article{Pedersen_PhysRevLett_2008,
  title = {Strong Pressure-Energy Correlations in van der Waals Liquids},
  author = {Pedersen, Ulf R. and Bailey, Nicholas P. and Schr\o{}der, Thomas B. and Dyre, Jeppe C.},
  journal = {Phys. Rev. Lett.},
  volume = {100},
  issue = {1},
  pages = {015701},
  numpages = {4},
  year = {2008},
  month = {Jan},
  publisher = {American Physical Society},
  doi = {10.1103/PhysRevLett.100.015701},
  url = {https://link.aps.org/doi/10.1103/PhysRevLett.100.015701}
}

@article{coslovich2009pressure,
  title={Pressure-energy correlations and thermodynamic scaling in viscous Lennard-Jones liquids},
  author={Coslovich, D and Roland, CM},
  journal={The Journal of Chemical Physics},
  volume={130},
  number={1},
  year={2009},
  publisher={AIP Publishing}
}

@article{Ninarello_PhysRevX_2017,
  title = {Models and Algorithms for the Next Generation of Glass Transition Studies},
  author = {Ninarello, Andrea and Berthier, Ludovic and Coslovich, Daniele},
  journal = {Phys. Rev. X},
  volume = {7},
  issue = {2},
  pages = {021039},
  numpages = {22},
  year = {2017},
  month = {Jun},
  publisher = {American Physical Society},
  doi = {10.1103/PhysRevX.7.021039}
}

@article{roland2005density,
  title={Density scaling of the dynamics of vitrifying liquids and its relationship to the dynamic crossover},
  author={Roland, CM and Casalini, R},
  journal={Journal of non-crystalline solids},
  volume={351},
  number={33-36},
  pages={2581--2587},
  year={2005},
  publisher={Elsevier}
}

@article{dyre2014hidden,
  title={Hidden scale invariance in condensed matter},
  author={Dyre, Jeppe C},
  journal={The Journal of Physical Chemistry B},
  volume={118},
  number={34},
  pages={10007--10024},
  year={2014},
  publisher={ACS Publications}
}

@article{veldhorst2014scaling,
  title={Scaling of the dynamics of flexible Lennard-Jones chains},
  author={Veldhorst, Arno A and Dyre, Jeppe C and Schr{\o}der, Thomas B},
  journal={The Journal of chemical physics},
  volume={141},
  number={5},
  year={2014},
  publisher={AIP Publishing}
}

@article{trachenko2021slow,
  title={Slow stretched-exponential and fast compressed-exponential relaxation from local event dynamics},
  author={Trachenko, Kostya and Zaccone, Alessio},
  journal={Journal of Physics: Condensed Matter},
  volume={33},
  number={31},
  pages={315101},
  year={2021},
  publisher={IOP Publishing}
}

@article{Berthier_PRX_2026,
  title = {Molecular Motion at the Experimental Glass Transition},
  author = {Simon, Romain and Barrat, Jean-Louis and Berthier, Ludovic},
  journal = {Phys. Rev. X},
  volume = {16},
  issue = {1},
  pages = {011035},
  numpages = {18},
  year = {2026},
  month = {Feb},
  publisher = {American Physical Society},
  doi = {10.1103/4twk-33j7}
}

@article{Pastore_2026,
  title = {Rare Cage Escapes Drive Relaxation in Deeply Supercooled Liquids},
  author = {Rusciano, Francesco and Pastore, Raffaele and Greco, Francesco and Kob, Walter},
  journal = {Phys. Rev. X},
  volume = {16},
  issue = {2},
  pages = {021061},
  numpages = {16},
  year = {2026},
  month = {Jun},
  publisher = {American Physical Society},
  doi = {10.1103/7m7x-zxqv},
  url = {https://link.aps.org/doi/10.1103/7m7x-zxqv}
}

@article{e_IS_2002,
  title = {Structural probe of a glass-forming liquid: Generalized compressibility},
  author = {Carruzzo, Herv\'e M. and Yu, Clare C.},
  journal = {Phys. Rev. E},
  volume = {66},
  issue = {2},
  pages = {021204},
  numpages = {16},
  year = {2002},
  month = {Aug},
  publisher = {American Physical Society},
  doi = {10.1103/PhysRevE.66.021204},
  url = {https://link.aps.org/doi/10.1103/PhysRevE.66.021204}
}
\end{document}